\documentclass[fleqn,10pt]{wlscirep}

\usepackage[utf8]{inputenc}
\usepackage[T1]{fontenc}
\usepackage{longtable}
\usepackage{float}
\usepackage{tabularx}

\title{Inter- and Intra-Subject Variability in EEG: A Systematic Survey}

\author[1,*]{Xuan-The Tran}
\author[2]{Thien-Nhan Vo}
\author[2,3]{Son-Tung Vu}
\author[2,4]{Thoa-Thi Tran}
\author[5]{Manh-Dat Nguyen}
\author[5]{Thomas Do}
\author[5]{Chin-Teng Lin}

\affil[1]{Vietnam Maritime University, Haiphong, Vietnam}
\affil[2]{HAI-Smartlink Research Lab, Anchi STE Company, Vietnam}
\affil[3]{Hanoi Architectural University, Hanoi, Vietnam}
\affil[4]{Greenwich Vietnam, FPT University, Hanoi, Vietnam}
\affil[5]{Computational Intelligence and Brain Computer Interface Lab, School of
Computer Science, Australian AI Institute, Faculty of Engineering and Information
Technology, University of Technology Sydney, Sydney, New South Wales, 2007,
Australia}

\affil[*]{thetx.vck@vimaru.edu.vn}

\keywords{electroencephalography, inter-subject variability, intra-subject
variability, test--retest reliability, brain--computer interface, literature review}

\begin{abstract}
Electroencephalography (EEG) underpins neuroscience, clinical neurophysiology, and brain--computer interfaces (BCIs), yet pronounced inter- and intra-subject variability limits reliability, reproducibility, and translation. This survey maps how that variability is conceptualised, quantified, and managed. From a documented search of six databases we screened 312 records and synthesised 87 studies spanning resting-state, event-related potential (ERP), and task-related/BCI paradigms in healthy and clinical cohorts. Four findings organise the field. First, stability is a property of the feature, not of EEG: eyes-closed posterior alpha power and individual alpha frequency are comparatively reliable, high-frequency power and single-edge connectivity estimates are not, and microstate topographies outlast their derived temporal parameters. Second, much observed variability is methodological, arising from electrode placement, referencing, pipeline choices, and evaluation design rather than from the brain. Third, paradigm and metric are confounded: motor-imagery work reports decoding accuracy while resting-state work reports intraclass correlations, so the two bodies of evidence rarely yield comparable numbers, and 31 of 64 paradigm~$\times$~metric combinations in our evidence map are unevidenced. Fourth, the evidence base is thinnest where robustness claims are strongest, with only two studies evaluating cross-dataset or multi-site designs. We map the biological, state-related, technical, and analytical sources of variability, compare quantification approaches (ICC, coefficient of variation, signal-to-noise ratio, generalizability theory, and pattern- and learning-based methods), and set out reporting recommendations. EEG variability is best treated as a design variable: a constraint to manage, and a structured signal to exploit for precision neuroscience and robust neurotechnology.
\end{abstract}

\begin{document}

\flushbottom
\maketitle
\thispagestyle{empty}

\section{Introduction}
Electroencephalography (EEG) is widely used for studying brain dynamics and for building applied neurotechnologies because it provides high temporal resolution and can be collected repeatedly across tasks and contexts. However, EEG measurements are strongly shaped by \emph{systems, subjects, and sessions}, and substantial variability is consistently observed both between individuals (inter-subject) and within the same individual across trials, days, and tasks (intra-subject)~\cite{melnik_systems_2017,saha_intra-_2020}. In practical decoding settings, this variability manifests as distribution shift across subjects, sessions, and datasets, directly degrading generalization and inflating performance estimates when evaluation protocols are not carefully designed~\cite{huang_discrepancy_2023,kamrud_effects_2021,xu_cross-dataset_2020}. Importantly, variability is multi-determined: it can reflect meaningful trait-like differences and state dynamics, but it is also amplified by technical and analytical factors (e.g., electrode placement variability, source\slash connectivity pipeline choices)~\cite{scrivener_variability_2022,allouch_effect_2023}.

Variability matters for at least four reasons. First, it is central to reliability and reproducibility: test--retest datasets and longitudinal designs show that EEG features can be stable in some conditions but change markedly with state and time, motivating explicit quantification of within- and across-session consistency~\cite{wang_test-retest_2022,meghdadi_inter_2024}. Second, variability constrains biomarker development and interpretation. Quantitative EEG (qEEG) biosignatures and effect-size interpretation depend on how inter- and intra-subject variability interact across frequencies and features, with implications for longitudinal change detection and clinical inference~\cite{meghdadi_inter_2024,liu_resting_2024,nahmias_consistency_2019}. 

Third, variability is a primary driver of BCI performance limitations: cross-session drift and cross-subject heterogeneity reduce decoding robustness, motivate recalibration, and shape the design of adaptation and transfer strategies; large multi-day and multi-session datasets have been introduced specifically to benchmark these effects~\cite{ma_large_2022,huang_discrepancy_2023,maswanganyi_statistical_2022}. Related work also links performance variation to physiological and state-related factors (e.g., pre-task and task-stage band power, stress), highlighting that variability can be partly predictable rather than purely random~\cite{zhou_relative_2021,zhang_stress-induced_2020,borgheai_multimodal_2024}. Finally, variability is also an opportunity for precision neuroscience: several studies indicate that individual signatures can be detectable and reproducible over time (e.g., trial-by-trial variability magnitudes; microstate and spectral ``fingerprints''), suggesting a trait-like component alongside state dependence~\cite{arazi_magnitude_2016,liu_reliability_2020,croce_eeg_2020,zulliger_within_2022}.

In this review, we use \textbf{inter-subject variability} to denote between-person differences in EEG features under comparable experimental contexts, and \textbf{intra-subject variability} to denote within-person fluctuations across trials, sessions, or tasks~\cite{saha_intra-_2020,huang_discrepancy_2023}. We distinguish variability from \textbf{reliability}, which concerns the consistency of measurements across repeated observations and is commonly assessed with test--retest designs and reliability metrics in qEEG and cognitive\slash rest datasets~\cite{wang_test-retest_2022,meghdadi_inter_2024,nahmias_consistency_2019}. A feature may show large absolute fluctuations but remain informative if individual structure is preserved; conversely, analytic flexibility can substantially alter outcomes (particularly for connectivity), emphasizing the need to interpret variability estimates in light of pipeline choices~\cite{allouch_effect_2023}.

We synthesize evidence from 87 empirical studies selected because they explicitly quantified or modeled EEG inter- and/or intra-subject variability across paradigms including resting-state, event-related potentials (ERPs), motor imagery (MI), steady-state visual evoked potentials (SSVEP), and other task-based protocols. The reviewed work spans diverse feature families (spectral power, ERP components, ERD/ERS indices, functional connectivity measures, microstate parameters, spatial patterns) and populations ranging from healthy cohorts to clinical groups, including studies explicitly targeting component- and paradigm-dependent data quality\slash variability in ERP measurements~\cite{wang_test-retest_2022,zhang_variations_2023,eyamu_prefrontal_2024}. We also include evidence from multi-subject, multi-session resources intended to characterize commonality and variability across tasks and days, supporting realistic generalization and benchmarking~\cite{huang_m3_2022,ma_large_2022}.

The corpus was assembled through a documented database search with explicit eligibility criteria, and the synthesis that follows is narrative rather than statistical~\cite{grant_typology_2009}. The design follows from the question. This literature does not pose ``what is the pooled effect'' --- the studies are too heterogeneous in paradigm, feature definition, retest interval, and outcome metric for pooling to be meaningful, and a motor-imagery decoding accuracy and a resting-state ICC are not commensurable quantities. It poses ``what has been examined, how, and what has not.'' We therefore report the search and selection transparently so that the corpus can be inspected, but we make no claim of compliance with a systematic-review reporting standard. The survey is systematic in the sense that its search and selection are documented and reproducible, but the review was not prospectively registered, screening was not independently duplicated, and no risk-of-bias appraisal was performed. Section~\ref{sec:limitations} states what this costs the reader.

The review is organised around four questions:
\begin{enumerate}
  \item[\textbf{RQ1}] How is EEG variability conceptualised, and which biological, state-related, technical, and analytical sources produce it? (Section~\ref{sec:conceptual_framework})
  \item[\textbf{RQ2}] Which metrics and models are used to quantify and represent variability, and what does each actually estimate? (Section~\ref{sec:methods})
  \item[\textbf{RQ3}] How large and how structured is variability across paradigms and feature families, and which features are reliable enough to support individual-level inference? (Section~\ref{sec:paradigms})
  \item[\textbf{RQ4}] What follows for study design, reporting, mitigation strategies, and translation to BCI, biomarker, and cognitive-neuroscience applications, and where is the evidence base empty? (Section~\ref{sec:discussion})
\end{enumerate}

The remainder of this article is organized as follows. Section~\ref{sec:methodology} describes the search methodology and the characteristics of the included studies. Sections~\ref{sec:conceptual_framework}--\ref{sec:discussion} address RQ1--RQ4 in turn, and Section~\ref{sec:conclusion} concludes. The complete evidence table for all 87 included studies is given in Appendix~A.

\section{Methodology}
\label{sec:methodology}

The review proceeded in four stages: defining the review questions, searching and screening for relevant studies, extracting a structured record from each included study, and synthesising the results narratively. The search and selection are documented and reproducible, while the synthesis is interpretive rather than statistical~\cite{grant_typology_2009}. We state plainly what was not done, because the distinction matters for how the findings should be weighed: no protocol was registered or published in advance, screening was carried out without an independently duplicated second pass or an inter-rater agreement statistic, and study quality was not formally appraised. Section~\ref{sec:limitations} returns to each of these.

The goal of the search was to capture research that either (i) directly quantified variability and reliability (e.g., test--retest stability, cross-session drift, cross-subject heterogeneity), or (ii) evaluated methodological strategies designed to manage, reduce, or exploit variability (e.g., normalization, harmonization, adaptation, calibration-efficient decoding).

Searches were conducted across six databases (Web of Science, PubMed, IEEE Xplore, ScienceDirect, SpringerLink, and Scopus), last run on 3 April 2025, and were supplemented by backward and forward citation screening of key papers. The query design combined terms for EEG with terms for variability and reliability, together with paradigm-specific terms to ensure coverage across resting-state, ERP, oscillatory, and BCI contexts. Representative keyword groups included: ``EEG'' OR ``electroencephalography'' AND (``variability'' OR ``inter-subject'' OR ``intra-subject'' OR ``within-subject'' OR ``between-subject'' OR ``test--retest'' OR ``reliability'' OR ``repeatability'' OR ``reproducibility'' OR ``non-stationarity'' OR ``session-to-session'' OR ``domain shift'' OR ``generalization'') with optional paradigm terms (e.g., ``resting-state'', ``ERP'', ``P300'', ``motor imagery'', ``SMR'', ``SSVEP'', ``BCI''). 
\subsection{Inclusion and Exclusion Criteria}
\begin{figure}[htbp]
    \centering
    \includegraphics[width=\textwidth]{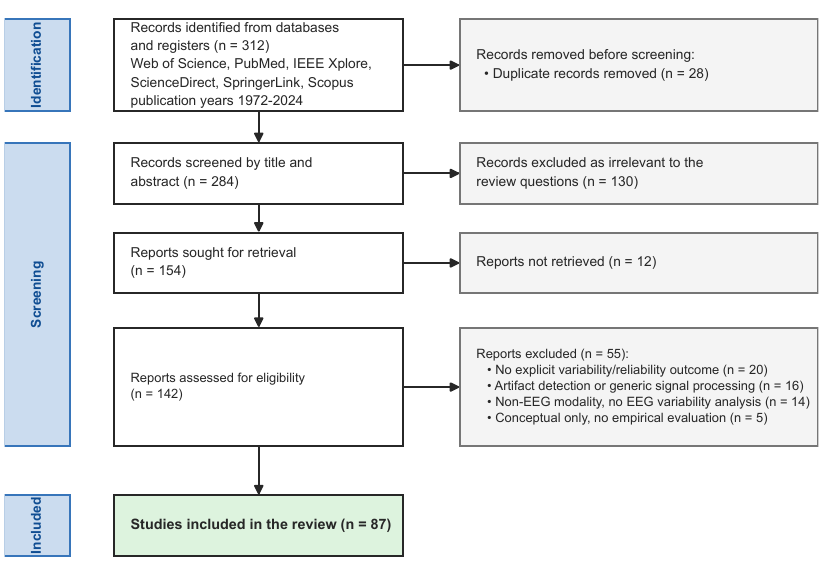}
    \caption{Search and selection flow. Of 312 records identified across six databases, 28 were duplicates (22 removed before screening and six found during data extraction), 130 records were excluded at title\slash abstract, and 12 could not be retrieved; of the 142 full texts then assessed, 55 were excluded for the reasons listed, leaving 87 studies. The four reasons restate the eligibility criteria given in the text.}
    \label{fig:selection}
\end{figure}

As summarized in Fig.~\ref{fig:selection}, records were identified through database search and then screened in multiple stages. An initial corpus of 312 records was assembled, duplicates were removed, and the remaining records were screened by title and abstract for relevance to the review questions. Full texts were then assessed for eligibility, resulting in 87 studies included in the final synthesis. Six duplicates were only identified during data extraction: two records had been entered twice, and four were preprints whose journal versions were also in the corpus. The published version of each was kept, and Fig.~\ref{fig:selection} counts the six among the duplicates. Included studies either (i) explicitly targeted EEG variability and/or reliability as a primary research question, or (ii) reported quantitative analyses of variability patterns (e.g., test--retest reliability, cross-session drift, cross-subject heterogeneity, cross-domain generalization) that were central to interpretation.

\paragraph{Eligibility criteria.}
Studies were included if they met one or more of the following criteria: they explicitly examined inter-subject and/or intra-subject variability in EEG (within-session, across-session, or across contexts); they reported quantitative estimates of variability or reliability (e.g., ICC, coefficient of variation, SNR-based metrics, pattern similarity, or performance generalization across subjects\slash sessions); they evaluated factors that modulate EEG variability (e.g., task, age, clinical status, recording protocol, montage\slash reference); or they proposed and empirically evaluated methodological approaches intended to manage, reduce, or exploit variability (e.g., harmonization, normalization, adaptation, calibration strategies). Records were excluded when they focused exclusively on artifact detection or generic signal-processing methodology without explicit variability analysis, addressed non-EEG modalities without direct EEG variability evaluation, or presented purely conceptual\slash methodological arguments without empirical variability or reliability outcomes.

\begin{figure}[htbp]
    \centering
    \includegraphics[width=\textwidth]{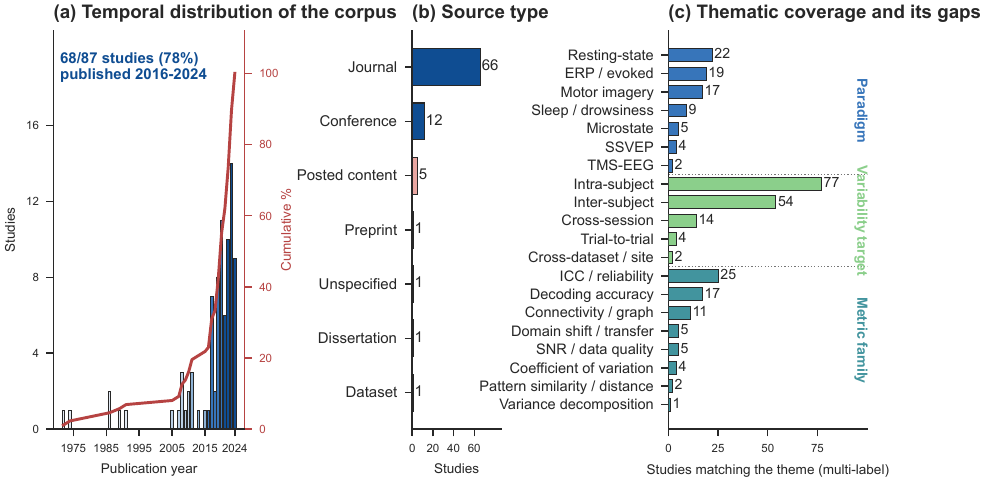}
    \caption{The evidence base is recent, largely peer-reviewed, and unevenly distributed across paradigms and metrics. (a) 68 of the 87 included studies (78\%) were published in 2016 or later; bars are colored by the reporting era and the red curve is the cumulative share. (b) 78 records are journal or conference articles; the remaining 9 are preprints, posted content, a dissertation and a dataset descriptor. (c) Thematic coding is multi-label, so a study contributes to every code it matches and column totals exceed 87; cross-dataset and multi-site evaluation (n=2) and formal variance decomposition (n=1) are the thinnest areas of the corpus.}
    \label{fig:corpus}
\end{figure}

\paragraph{Temporal distribution of the evidence base.}
The included studies span more than five decades (1972--2024) and show a pronounced increase in publication volume in the last decade (Fig.~\ref{fig:corpus}a). The distribution across time periods was: 1970s--1990s (6 studies), 2000--2010 (8 studies), 2011--2015 (5 studies), 2016--2020 (29 studies), and 2021--2024 (39 studies), summing to the 87 included studies. Sixty-eight of the 87 studies (78\%) were published in 2016 or later. This growth is consistent with (i) increased emphasis on reproducibility and reliability reporting, (ii) expanding use of EEG for individualized prediction and biomarker development, and (iii) practical demand for robust, calibration-efficient BCIs and generalizable machine learning models. Recent studies more frequently evaluate variability under ecologically realistic conditions (multi-session, multi-site, cross-dataset) and increasingly frame variability as a domain-shift problem rather than solely measurement noise.

\subsection{Screening and Data Extraction}
Screening proceeded in three stages: (1) title\slash abstract screening to remove clearly irrelevant records; (2) full-text eligibility assessment against inclusion\slash exclusion criteria; and (3) structured data extraction from the included studies.

We did not appraise the methodological quality or risk of bias of the included studies. This is a deliberate choice rather than an oversight: the review's purpose is to map what has been measured and how, and a quality score computed across designs as different as a two-session ICC study and a cross-dataset deep-learning benchmark would not be interpretable on a common scale. It does mean that the findings below should be read as a description of the literature, not as a graded summary of evidence strength.

\paragraph{Reference list and reviewed corpus.}
The reference list is longer than the reviewed corpus, and the two should not be confused. It contains the 87 reviewed studies plus a smaller set of further works cited for context that never entered the screening process: one methodological source for the review design itself~\cite{grant_typology_2009}, and background references used to situate specific claims. Only the 87 reviewed studies are counted in Fig.~\ref{fig:selection}, tallied in Fig.~\ref{fig:corpus}, and listed in Appendix~A; every quantitative statement about the corpus refers to those 87 alone. For each included study, we extracted and coded (at minimum) population characteristics, paradigm\slash task, recording context (single-session vs.\ multi-session; single-site vs.\ multi-site), feature family (spectral/ERP/connectivity/microstates/BCI decoding), variability targets (inter-subject, intra-subject, or both), and the reported variability\slash reliability metrics (e.g., ICC specification when applicable, CV, correlation-based stability, variance components, or generalization performance). Studies were then grouped for descriptive synthesis by paradigm and feature family to enable comparison of reliability ranges under broadly similar methodological conditions, and to support later sections that discuss how variability sources and mitigation strategies differ across application domains.

Figure~\ref{fig:corpus} summarizes the resulting corpus. Sixty-six of the 87 records are journal articles and 12 are conference proceedings; the remaining 9 are preprints, posted content, a dissertation and a dataset descriptor, which we retained because several are the primary description of multi-session resources that the variability literature depends on, but which we flag here because they have not undergone peer review (Fig.~\ref{fig:corpus}b).

Cohort and paradigm coding is \emph{multi-label}: a study that evaluates two paradigms, or that reports both inter- and intra-subject outcomes, contributes to every applicable category, so category totals exceed 87 and should not be read as a partition. Coded this way, the corpus is dominated by healthy adult cohorts, with smaller groups of clinical cohorts spanning neurological and psychiatric disorders and of age-stratified (elderly and pediatric/adolescent) cohorts; the remaining studies did not report a cohort description precise enough to code. By paradigm, resting-state recordings (n=22), ERP/evoked paradigms (n=19) and motor imagery (n=17) are the best represented, followed by sleep/vigilance protocols (n=9), microstate analyses (n=5), SSVEP (n=4) and TMS--EEG (n=2); a further 29 studies are cross-paradigm or purely methodological and are not attributable to a single canonical paradigm. By variability target, 77 studies address intra-subject variation and 54 address inter-subject variation, with 14 explicitly evaluating cross-session and only 2 evaluating cross-dataset or multi-site designs (Fig.~\ref{fig:corpus}c). This last figure is itself a finding: the shift toward treating variability as a domain-shift problem, which the recent literature advocates, is not yet matched by the evidence base.

\section{Conceptual Framework of EEG Variability}
\label{sec:conceptual_framework}

\subsection{Sources of Variability}

\begin{figure}[htbp]
\centering
\includegraphics[width=\textwidth]{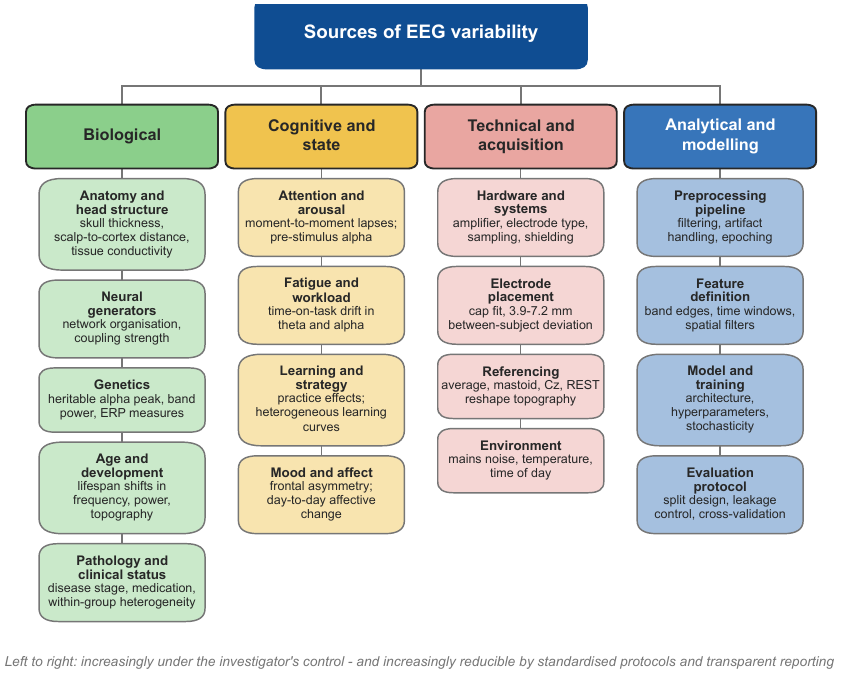}
\caption{EEG variability is produced by four interacting families of sources, only the right-hand two of which an investigator can directly control. Biological and state-related sources are properties of the participant and must be modeled or stratified; technical and analytical sources are properties of the protocol and pipeline and can be reduced by standardization and transparent reporting. Each column carries one color from header to leaf; no relation is asserted between leaves of the same column.}
\label{fig:sources_variability}
\end{figure}

EEG variability arises from a complex interplay of biological, psychological, technical, and analytical factors. Disentangling these sources is essential for interpreting observed effects, designing reproducible experiments, and developing models that are robust to non-stationarity and domain shift~\cite{huang_discrepancy_2023,melnik_systems_2017}. Importantly, the same observed variability can have different origins depending on the context: for example, changes across sessions may reflect genuine neurophysiological dynamics, but may also be driven by electrode placement differences, impedance fluctuations, or preprocessing choices. As illustrated in Fig.~\ref{fig:sources_variability}, we organize these major sources of variability into four broad categories and highlight their typical manifestations below.

\subsubsection*{Biological sources}
\textbf{Anatomy and head--brain structure.} Individual differences in skull thickness, scalp-to-cortex distance, tissue conductivity, and cortical folding influence both the amplitude and spatial distribution of scalp-recorded potentials~\cite{scrivener_variability_2022,bodmer_neurophysiological_2017}. These factors systematically contribute to inter-subject differences in feature magnitude and topography, and they can also modulate how sensitive a given montage is to specific cortical generators~\cite{scrivener_variability_2022}. Forward modeling and simulation studies indicate that realistic variation in head geometry and conductivity can yield substantial differences in recorded amplitudes and spatial patterns even when underlying neural sources are identical~\cite{bodmer_neurophysiological_2017,scrivener_variability_2022}. Such effects are especially relevant when comparing topographic biomarkers, source-localized estimates, or connectivity measures that depend on spatial covariance structure.

\textbf{Neural generators and functional organization.} Individuals differ in the organization and dynamics of functional brain networks, including the coupling strength and configuration of large-scale interactions~\cite{pani_subject_2019,nakuci_within-subject_2023}. These differences contribute to inter-subject variability in EEG connectivity and graph-theoretic measures~\cite{allouch_effect_2023}. Resting-state network properties can exhibit person-specific structure that is partly stable over time, consistent with the notion that some connectivity signatures behave as trait-like markers~\cite{nakuci_within-subject_2023}.

\textbf{Genetic influences.} Twin and family studies support a substantial heritable component in several EEG features, including alpha peak frequency, band-limited power, and selected ERP measures. Reported heritability estimates vary by feature and paradigm but can be sizable, implying that inter-subject variability includes stable biological contributions beyond measurement noise. Candidate genetic variants, particularly those related to neurotransmitter systems, have been associated with variability in oscillatory and evoked responses, although effect sizes and reproducibility can differ across cohorts and analytic choices.

\textbf{Age, development, and aging.} Developmental processes and aging-related neurophysiological changes are major drivers of both inter- and intra-subject variability~\cite{dong_modeling_2024,magnuson_increased_2020}. Across the lifespan, EEG spectral profiles and evoked responses show systematic shifts in frequency, power, and topography~\cite{milne_increased_2011,bodmer_neurophysiological_2017}. Interestingly, developmental studies show that brain signal variability increases with age and maturation, which paradoxically accompanies a decrease in behavioral response variability, indicating improved neural network optimization~\cite{mcintosh2008increased}. For example, alpha peak frequency tends to increase through childhood and adolescence and may decrease in older adulthood, while ERP component amplitudes and latencies change with maturation and cognitive aging~\cite{bodmer_neurophysiological_2017,magnuson_increased_2020}. These effects emphasize the importance of age matching, age-stratified analyses, and developmental considerations in biomarker and BCI studies.

\textbf{Pathology and clinical status.} Neurological and psychiatric disorders can alter mean EEG features and often increase within-group heterogeneity~\cite{liu_resting_2024,eyamu_prefrontal_2024}. Elevated variability in clinical cohorts may reflect heterogeneity in disease mechanisms, stage, symptom severity, medication status, and compensatory strategies~\cite{milne_increased_2011,magnuson_increased_2020}. For instance, longitudinal studies on infant cohorts demonstrate that developmental trajectories of EEG power during the first year can distinguish later autism spectrum disorder (ASD) outcomes~\cite{gabarddurnam2019longitudinal}, and day-to-day test-retest reliability of resting EEG spectral profiles remains high in children with ASD, validating their potential as stable diagnostic biomarkers~\cite{levin2020daytoday}. Consequently, variability itself can be informative (e.g., indicating subtypes), but it can also dilute group-level contrasts if not modeled appropriately.

\subsubsection*{Cognitive and psychological state sources}
\textbf{Attention and arousal.} Moment-to-moment fluctuations in attention and arousal are key drivers of within-session variability~\cite{arazi_magnitude_2016}. Lapses in attention can change ongoing oscillatory activity and modulate evoked response amplitudes and latencies, influencing both spectral features and ERP measures~\cite{arazi_magnitude_2016}. Pre-stimulus markers (e.g., alpha power) often correlate with subsequent perceptual performance and the magnitude of stimulus-evoked responses, linking intra-subject variability to latent state dynamics~\cite{arazi_magnitude_2016}.

\textbf{Fatigue and mental workload.} Prolonged tasks and sustained interaction with BCI systems can induce fatigue that systematically shifts EEG characteristics over time~\cite{hwang_mitigating_2021}. Fatigue-related changes are frequently reflected in increased low-frequency power (e.g., theta), altered alpha dynamics, and reduced amplitudes of cognitive ERP components, which may degrade decoding performance and confound longitudinal comparisons~\cite{hwang_mitigating_2021}.

\textbf{Learning, practice, and strategy adaptation.} Repeated exposure to tasks can elicit learning-dependent changes in EEG patterns across sessions~\cite{huang_discrepancy_2023}. In BCI paradigms, improvements in control can correspond to the acquisition of stable neural strategies, representing intra-subject variability that is meaningful rather than undesirable~\cite{huang_discrepancy_2023}. However, learning trajectories differ substantially across individuals, thereby contributing to inter-subject variability in calibration requirements and achievable performance~\cite{saha_can_2023}.

\textbf{Mood and affective state.} Variations in mood and emotion can influence EEG, particularly measures involving frontal activity and asymmetry indices. Day-to-day changes in affect may contribute to test--retest variability and can confound longitudinal designs unless measured and modeled as covariates.

\subsubsection*{Technical and acquisition sources}
\textbf{Hardware and recording systems.} EEG systems differ in channel count, electrode type (wet, dry, active), amplifier characteristics, dynamic range, sampling rate, and shielding, leading to cross-system variability even when measuring identical tasks~\cite{melnik_systems_2017}. Such differences contribute to cross-lab and cross-dataset heterogeneity and can interact with preprocessing choices (e.g., filtering)~\cite{melnik_systems_2017}. Even within a fixed system, impedance fluctuations and subtle placement differences can induce substantial variability in recorded amplitude and noise characteristics~\cite{scrivener_variability_2022}.

\textbf{Electrode placement and referencing.} Small deviations in electrode placement across sessions (despite standardized layouts such as the 10--20 system) can change the spatial sampling of cortical sources and alter feature estimates~\cite{scrivener_variability_2022}. Reference choice (e.g., average reference, linked mastoids, Cz, REST) can substantially reshape scalp topographies and influence connectivity measures via changes in covariance structure~\cite{allouch_effect_2023}. Inconsistent referencing across studies is therefore a major contributor to cross-dataset variability and complicates meta-analytic synthesis~\cite{allouch_effect_2023}.

\textbf{Environmental conditions.} Ambient electrical noise (e.g., mains interference), temperature, humidity, and local electromagnetic environments influence recording quality and artifact prevalence. Additionally, time of day and circadian rhythms can modulate EEG dynamics, including systematic variations in alpha frequency and power, thereby contributing to within- and across-session variability when protocols are not time-controlled.

\subsubsection*{Analytical and modeling sources}
\textbf{Preprocessing pipelines.} Filtering choices, artifact handling (manual rejection, ICA-based correction, automated pipelines), re-referencing, and segmentation parameters can substantially alter derived EEG measures~\cite{allouch_effect_2023,zhang_variations_2023}. Differences in pipelines across groups constitute a major source of cross-lab heterogeneity, and some evidence suggests that preprocessing decisions can affect not only mean feature values but also reliability estimates and apparent stability across sessions~\cite{zhang_variations_2023}. For variability-focused studies, it is therefore important to report preprocessing details comprehensively and, where feasible, evaluate sensitivity to plausible pipeline variants.

\textbf{Feature definitions and extraction.} The choice of feature family (e.g., band power vs.\ phase-based metrics vs.\ connectivity), frequency band boundaries, time windows, spatial filtering (e.g., CSP, beamformers), and normalization strategies shapes both the magnitude of observed variability and the resulting reliability~\cite{zhang_variations_2023}. For example, individualized frequency definitions (e.g., using individual alpha frequency) may reduce spurious inter-subject variance while preserving meaningful differences, though such normalization can also change interpretability and comparability across studies.

\textbf{Model architecture, training, and evaluation.} In learning-based pipelines, model class, hyperparameter selection, optimization stochasticity, and data augmentation can produce variability in performance metrics that is distinct from neurophysiological variability~\cite{kamrud_effects_2021}. Evaluation protocols also matter: cross-validation design, train--test splitting strategy, and leakage control can inflate or deflate apparent generalization and thereby alter conclusions about inter- and intra-subject robustness~\cite{kamrud_effects_2021}. Variability-aware reporting should therefore separate (i) variability intrinsic to EEG generation and measurement from (ii) variability induced by modeling and evaluation procedures.

\subsection{Relationship to Reliability and Test--Retest}
Variability and reliability are closely related but conceptually distinct. Variability describes the \emph{magnitude and structure} of dispersion (within a person, between people, across contexts), whereas reliability concerns the \emph{repeatability} of a measurement (that is, the extent to which observed differences are reproducible under repeated assessment)~\cite{meghdadi_inter_2024}. Clarifying this distinction is essential for interpreting EEG findings, selecting features for biomarkers or BCI control, and designing studies that can separate true neurophysiological change from measurement noise.

\paragraph{Distinguishing variability from reliability.}
High variability does not necessarily imply low reliability. For example, an EEG feature may exhibit large inter-subject differences (high between-person variance) but remain highly reliable if individuals maintain a consistent rank ordering across sessions~\cite{arazi_magnitude_2016}. Conversely, a feature may show low apparent variability at the group level but still be unreliable if repeated measurements fluctuate unpredictably within individuals or are strongly affected by uncontrolled state or technical factors~\cite{liu_resting_2024}. In practice, reliability depends on the relative contributions of between-subject variance, within-subject variance, and measurement error; thus, the same absolute variability can correspond to different reliability profiles depending on the study design and the population sampled.

\section{Methods to Quantify and Model EEG Variability}
\label{sec:methods}

\begin{figure}[htbp]
\centering
\includegraphics[width=\textwidth]{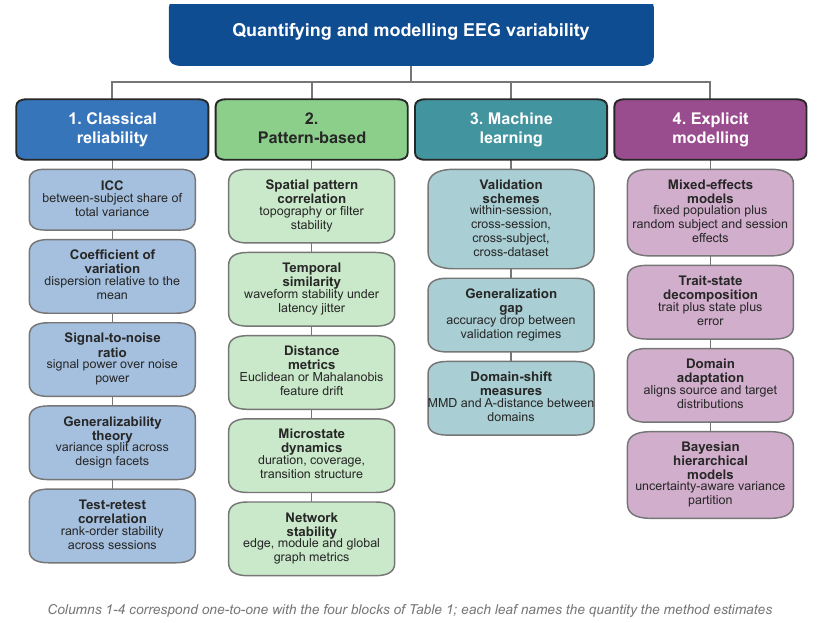}
\caption{Four families of methods answer different questions about the same data: classical indices summarize stability of a scalar feature, pattern-based measures summarize stability of a whole spatial or temporal structure, learning-based schemes express variability as a generalization gap, and explicit models partition variance rather than discarding it. Columns 1--4 correspond one-to-one with the four blocks of Table~\ref{tab:eeg_variability}; each leaf names the quantity the method estimates.}
\label{fig:EEGreliabilityvariability}
\end{figure}

\begin{table}
\centering
\caption{Summary of Methods to Quantify and Model EEG Variability}
\label{tab:eeg_variability}
\scriptsize
\setlength{\tabcolsep}{3.5pt}
\renewcommand{\arraystretch}{1.06}
\resizebox{\textwidth}{!}{%
\begin{tabular}{p{2.1cm} p{2.6cm} p{3.0cm} p{5.6cm} p{2.8cm} p{4.2cm}}
\toprule
\textbf{Category} & \textbf{Method} & \textbf{Study (examples)} & \textbf{Key Concept / Formula} & \textbf{Application} & \textbf{Limitations} \\
\midrule
\textbf{1.Classical Reliability}
& ICC
& \cite{maswanganyi_statistical_2022,ma_large_2022,meghdadi_inter_2024,saha_intra-_2020}
& Ratio: $\frac{\sigma^2_{between}}{\sigma^2_{total}}$
& Test--retest reliability quantification.
& Sensitive to sample homogeneity \& ICC model choice. \\
\addlinespace
& Coefficient of Variation (CV)
& \cite{meghdadi_inter_2024,wang_test-retest_2022,maswanganyi_statistical_2022,saha_intra-_2020}
& $\frac{\sigma}{\mu}\times 100\%$
& Comparing dispersion across measures\slash scales.
& Dispersion, not reproducibility; unstable when $\mu \approx 0$. \\
\addlinespace
& Signal-to-Noise Ratio (SNR)
& \cite{zhang_variations_2023,wang_test-retest_2022}
& Ratio: $P_{signal}/P_{noise}$
& ERP/spectral data quality assessment.
& High SNR $\nRightarrow$ high reliability (e.g., systematic shifts). \\
\addlinespace
& Generalizability Theory (G-Theory)
& \cite{maswanganyi_statistical_2022,eckert_statistical_1974}
& Variance decomposition across facets (subj, sess, item, \dots)
& Designing studies (e.g., trial\slash session counts) and partitioning sources.
& Needs sufficient\slash structured data; can be complex to implement\slash interpret. \\
\addlinespace
& Test--retest correlation
& \cite{wang_test-retest_2022,meghdadi_inter_2024,nahmias_consistency_2019}
& Pearson\slash Spearman $r$ across sessions
& Rank-order stability check.
& Ignores mean shifts\slash absolute agreement; outlier sensitive. \\
\midrule
\textbf{2.Pattern Approaches}
& Pattern similarity
& \cite{wang_test-retest_2022,maswanganyi_statistical_2022,zhang_variations_2023}
& Correlation of vectors/waveforms/maps across time
& Stability of spatial patterns or ERP morphology\slash timing.
& Affected by electrode shifts, referencing, and latency jitter. \\
\addlinespace
& Distance metrics
& \cite{wang_test-retest_2022,huang_discrepancy_2023}
& Euclidean\slash Mahalanobis (feature-space distance)
& Drift detection; cross-session feature stability in decoding.
& Depends on feature scaling\slash preprocessing; metric choice matters. \\
\addlinespace
& Microstate dynamics
& \cite{saha_intra-_2020,zanesco_within_2019,liu_reliability_2020}
& Stability of microstate parameters / transition structure
& Whole-brain state sequence characterization.
& Temporal parameters can be less stable than spatial topographies. \\
\addlinespace
& Network\slash connectivity stability
& \cite{saha_intra-_2020,pani_subject_2019,nakuci_within-subject_2023,allouch_effect_2023}
& Reproducibility of graph edges/metrics/modules
& Functional connectivity organization across sessions\slash subjects.
& Single edges often noisy; pipeline choices can dominate outcomes. \\
\midrule
\textbf{3.ML Perspectives}
& Cross-validation (WS/CS/CV)
& \cite{ma_large_2022,kamrud_effects_2021}
& Performance under within-session vs.\ cross-session\slash subject splits
& Operationalizing variability as generalization gap.
& Sensitive to partitioning; improper splits can inflate accuracy. \\
\addlinespace
& Domain shift quantification
& \cite{huang_discrepancy_2023,xu_cross-dataset_2020}
& Distribution mismatch (e.g., MMD, classifier-based distances)
& Measuring train--test mismatch across sessions\slash datasets.
& Requires careful estimation; depends on feature representation. \\
\midrule
\textbf{4.Explicit Modeling}
& Mixed-effects models
& \cite{eckert_statistical_1974,dong_modeling_2024}
& Fixed (population) + random (subject\slash session) effects
& Unbalanced longitudinal data; individual trajectories.
& Complexity grows with interactions; modeling assumptions matter. \\
\addlinespace
& Trait--state decomposition
& \cite{saha_intra-_2020,zanesco_within_2019}
& $Y = Trait + State + Error$
& Separating stable traits from fluctuating states.
& Requires repeated measures and state\slash context annotation. \\
\addlinespace
& Domain adaptation / transfer learning
& \cite{huang_discrepancy_2023,xu_cross-dataset_2020}
& Aligning source\slash target feature distributions
& Mitigating cross-session\slash cross-subject performance drops.
& May require target calibration data; risk of negative transfer. \\
\addlinespace
& Bayesian hierarchical models
& \cite{wang_test-retest_2022,nakuci_within-subject_2023}
& Probabilistic multi-level modeling with uncertainty
& Variance partitioning and uncertainty-aware inference.
& Computationally intensive; requires careful priors\slash diagnostics. \\
\bottomrule
\end{tabular}%
}
\end{table}

A rigorous understanding of inter- and intra-subject variability in EEG depends on methodological choices that determine \emph{what} is treated as signal, \emph{how} variability is quantified, and \emph{which} sources of variation are modeled explicitly versus absorbed into error. As illustrated in the methodological overview (Fig.~\ref{fig:EEGreliabilityvariability}), these approaches span classical reliability indices, pattern-based similarity, and learning-based cross-validation. This section reviews common methods used in the literature to quantify and model variability (summarized in Table~\ref{tab:eeg_variability}), and to formalize variability structure in ways that support study design, longitudinal inference, and robust decoding.

\subsection{Classical Reliability Metrics}
\label{sec:classical_metrics}
Classical reliability metrics remain widely used because they provide interpretable summaries of stability across repeated measurements ~\cite{meghdadi_inter_2024}. However, each metric captures a distinct aspect of variability, and none is sufficient on its own for characterizing the full spatiotemporal and contextual complexity of EEG. Accordingly, variability-focused EEG studies increasingly report multiple indices and pair them with explicit modeling of design factors (e.g., session interval, trial count, montage, preprocessing pipeline) ~\cite{maswanganyi_statistical_2022} ~\cite{wang_test-retest_2022}.

\paragraph{Intraclass correlation coefficient (ICC).}
The intraclass correlation coefficient is the most commonly used statistic for quantifying test--retest reliability in EEG research~\cite{meghdadi_inter_2024}. Conceptually, ICC expresses the fraction of total observed variance attributable to stable between-subject differences relative to within-subject variance (including measurement error and state-driven fluctuation). In a simplified two-level setting, ICC can be expressed as:
\begin{equation}
\mathrm{ICC} \;=\; \frac{\sigma^2_{\mathrm{between}}}{\sigma^2_{\mathrm{between}} + \sigma^2_{\mathrm{within}}},
\end{equation}
where $\sigma^2_{\mathrm{between}}$ is between-subject variance and $\sigma^2_{\mathrm{within}}$ is within-subject variance across repeated assessments.

\textbf{Interpretation.} Conventional heuristic thresholds are often used to describe ICC magnitudes (e.g., poor, fair, good, excellent), such as:
\begin{itemize}
    \item $\mathrm{ICC}<0.25$: poor,
    \item $0.25\le \mathrm{ICC}<0.50$: fair,
    \item $0.50\le \mathrm{ICC}<0.75$: good,
    \item $\mathrm{ICC}\ge 0.75$: excellent ~\cite{meghdadi_inter_2024}.
\end{itemize}
These thresholds should be interpreted in context because acceptable reliability depends on the intended application (e.g., group-level inference vs.\ individual-level clinical decisions vs.\ longitudinal change detection) and on the costs of misclassification or measurement error~\cite{meghdadi_inter_2024}.

\textbf{ICC variants and design dependence.} Multiple ICC formulations exist, reflecting whether raters\slash sessions are treated as random or fixed effects and whether single measurements or averages are considered. Common forms include ICC(1,1), ICC(2,1), and ICC(3,1)~\cite{meghdadi_inter_2024}. In many test--retest EEG contexts, ICC(2,1) or ICC(3,1) are used depending on whether the specific sessions\slash raters are considered a random sample from a larger universe (random effects) or are the only conditions of interest (fixed effects)~\cite{meghdadi_inter_2024,maswanganyi_statistical_2022}. Because ICC estimates can differ meaningfully across formulations, variability-focused work should report the exact ICC model, the unit of analysis (single trial, condition average, session average), and confidence intervals where possible.

\textbf{Limitations.} ICC is informative but has important caveats:
\begin{itemize}
    \item \emph{Dependence on between-subject variance:} low ICC can arise either from high within-subject variability or from low between-subject heterogeneity (e.g., a homogeneous sample), even when measurement error is small~\cite{wang_test-retest_2022,ma_large_2022,saha_intra-_2020}.
    \item \emph{Model assumptions:} standard ICC formulations assume linear relationships and homoscedasticity, which may be violated for EEG features with non-Gaussian distributions or state-dependent variance~\cite{maswanganyi_statistical_2022,ma_large_2022}.
    \item \emph{Sample size sensitivity:} ICC can be unstable in small samples, especially when estimating multiple variance components or when outliers are present~\cite{maswanganyi_statistical_2022,wang_test-retest_2022}.
\end{itemize}

\paragraph{Coefficient of variation (CV).}
The coefficient of variation expresses variability relative to the mean:
\begin{equation}
\mathrm{CV} \;=\; \left(\frac{\sigma}{\mu}\right)\times 100\%,
\end{equation}
where $\sigma$ is the standard deviation and $\mu$ is the mean of the measure.

\textbf{Applications.} CV is frequently used to:
\begin{itemize}
    \item compare relative dispersion of measures expressed on different scales~\cite{maswanganyi_statistical_2022},
    \item summarize measurement precision and heteroscedasticity patterns~\cite{maswanganyi_statistical_2022,ma_large_2022},
    \item flag unusually variable or potentially artifactual measurements~\cite{ma_large_2022,saha_intra-_2020}.
\end{itemize}

\textbf{Limitations.} CV does not directly quantify reliability because it does not assess whether individual differences are reproducible across sessions. A high CV may reflect unstable measurement, genuine heterogeneity across individuals, or a combination of both; thus, CV should be interpreted alongside reliability indices such as ICC, test--retest correlation, or variance-component\slash trait--state frameworks~\cite{meghdadi_inter_2024,wang_test-retest_2022,maswanganyi_statistical_2022,saha_intra-_2020}.

\paragraph{Signal-to-noise ratio (SNR).}
SNR compares the magnitude of a signal of interest to background noise. In power terms:
\begin{equation}
\mathrm{SNR} \;=\; \frac{P_{\mathrm{signal}}}{P_{\mathrm{noise}}},
\end{equation}
and in decibels:
\begin{equation}
\mathrm{SNR}_{\mathrm{dB}} \;=\; 10\log_{10}\left(\frac{P_{\mathrm{signal}}}{P_{\mathrm{noise}}}\right).
\end{equation}

\textbf{Applications in EEG.} SNR is used across multiple EEG domains:
\begin{itemize}
    \item \textbf{ERP research:} SNR may be quantified as the ratio of averaged ERP amplitude to baseline variability (or other noise estimates), and is closely tied to single-trial variability and measurement error considerations~\cite{zhang_variations_2023,milne_increased_2011,magnuson_increased_2020,ganin_sources_2023}.
    \item \textbf{Spectral analysis:} SNR can be defined as power in a target band relative to adjacent bands or noise floors, often used for oscillatory peaks and band-limited contrasts~\cite{saha_intra-_2020,meghdadi_inter_2024,zhou_relative_2021}.
    \item \textbf{BCI:} feature-level SNR (e.g., discriminative band-power contrasts or ERD/ERS strength) is often predictive of decoding performance and can be used as a proxy for expected classifier separability and user ``BCI efficiency''~\cite{zhou_relative_2021,huang_discrepancy_2023,zhang_stress-induced_2020}.
\end{itemize}

\textbf{Relationship to reliability.} Higher SNR often, though not invariably, corresponds to improved reliability because noise-driven fluctuations contribute less to the measured feature~\cite{zhang_variations_2023,meghdadi_inter_2024,wang_test-retest_2022}. However, reliability can still be low when systematic shifts occur across sessions (e.g., electrode placement changes, state drift, learning effects), even if within-session SNR is high~\cite{wang_test-retest_2022,ma_large_2022,scrivener_variability_2022,ribeiro_slow_2021}. Consequently, SNR is best interpreted as a necessary but not sufficient condition for reliability.

\paragraph{Generalizability theory (G-theory).}
Generalizability theory extends classical reliability by decomposing variability into multiple facets (e.g., subject, session, item\slash stimulus, rater, and their interactions)~\cite{maswanganyi_statistical_2022,eckert_statistical_1974}. A generic variance decomposition can be written as:
\begin{equation}
\sigma^2_{\mathrm{total}} \;=\; \sigma^2_{\mathrm{subjects}} + \sigma^2_{\mathrm{sessions}} + \sigma^2_{\mathrm{items}}
+ \sigma^2_{\mathrm{interactions}} + \sigma^2_{\mathrm{error}}.
\end{equation}

\textbf{Advantages.} G-theory provides:
\begin{itemize}
    \item simultaneous estimation of multiple sources of variability, rather than collapsing them into a single error term~\cite{maswanganyi_statistical_2022,eckert_statistical_1974};
    \item a principled way to optimize design (e.g., how many trials or sessions are needed to achieve a target reliability), closely related to work emphasizing trial counts, measurement error, and data quality in ERP/EEG~\cite{zhang_variations_2023,wang_test-retest_2022};
    \item generalizability coefficients that are conceptually analogous to reliability coefficients but explicitly design-dependent~\cite{maswanganyi_statistical_2022}.
\end{itemize}

\textbf{EEG applications.} In the included literature, G-theory has been used to motivate variance partitioning across subjects, sessions, and task\slash stimulus facets when interpreting stability and reproducibility of EEG-derived measures~\cite{maswanganyi_statistical_2022,wang_test-retest_2022,zhang_variations_2023}. Related multi-facet perspectives also arise in work emphasizing that systems, subjects, and sessions jointly shape EEG outcomes and apparent reliability~\cite{melnik_systems_2017}.

\textbf{Limitations.} G-theory can require balanced or well-structured designs and becomes computationally and statistically complex when many facets are included or when data are highly unbalanced (common in real-world EEG due to artifact-related trial loss)~\cite{maswanganyi_statistical_2022,zhang_variations_2023}.

\paragraph{Test--retest correlation.}
Simple Pearson or Spearman correlations between measurements at two time points provide an intuitive assessment of rank-order stability:
\begin{equation}
r \;=\; \mathrm{corr}\left(X_{\mathrm{Time1}}, X_{\mathrm{Time2}}\right).
\end{equation}

\textbf{Advantages.} Correlation is easy to compute and interpret and directly assesses whether individuals maintain relative ordering across sessions~\cite{wang_test-retest_2022,meghdadi_inter_2024,nahmias_consistency_2019}.

\textbf{Limitations.} Correlation-based indices have several drawbacks:
\begin{itemize}
    \item they do not account for systematic mean shifts across sessions (e.g., practice effects, habituation), and thus can indicate high stability even when absolute agreement is poor~\cite{wang_test-retest_2022,meghdadi_inter_2024};
    \item they can be sensitive to outliers, especially in small samples and in the presence of data-quality heterogeneity across participants~\cite{maswanganyi_statistical_2022,zhang_variations_2023,wang_test-retest_2022};
    \item they do not explicitly separate between-subject from within-subject variance components, limiting interpretability when the goal is to model sources of variability (e.g., subject vs session vs trial factors)~\cite{maswanganyi_statistical_2022,eckert_statistical_1974,melnik_systems_2017}.
\end{itemize}
For these reasons, test--retest correlations are often best reported alongside ICC and/or variance-component approaches, particularly in studies aiming to support individual-level inference or longitudinal change detection~\cite{meghdadi_inter_2024,maswanganyi_statistical_2022}.

\subsection{Multivariate and Pattern-Based Approaches}
Classical reliability statistics (e.g., ICC, test--retest correlation) are typically computed on \emph{scalar} features such as band power at a given electrode or ERP amplitude in a fixed time window. However, many modern EEG analyses and BCI pipelines rely on \emph{multivariate} representations (topographies, spatial filters, covariance matrices, time--frequency maps, connectivity graphs), where the object of interest is the \emph{pattern} rather than any single element~\cite{saha_intra-_2020,melnik_systems_2017,pani_subject_2019,nakuci_within-subject_2023}. Pattern-based approaches therefore quantify stability by measuring similarity (or dissimilarity) between high-dimensional representations obtained from different trials, sessions, or datasets~\cite{wang_test-retest_2022,saha_intra-_2020,maswanganyi_statistical_2022,huang_discrepancy_2023,xu_cross-dataset_2020}.

\subsubsection*{Pattern similarity metrics}
\textbf{Spatial pattern correlation.} For scalp topographies, spatial filters (e.g., CSP weights), or source maps, a common approach is to compute correlation between vectors representing the pattern across electrodes:
\begin{equation}
r_{\mathrm{spatial}} \;=\; \mathrm{corr}\!\left(\mathbf{p}^{(1)},\,\mathbf{p}^{(2)}\right),
\end{equation}
where $\mathbf{p}^{(1)}$ and $\mathbf{p}^{(2)}$ denote the spatial patterns from session 1 and session 2, respectively. High spatial correlations are commonly interpreted as evidence of stable topography~\cite{wang_test-retest_2022,maswanganyi_statistical_2022}. This strategy has been used in reliability-oriented analyses to assess stability of ERP-related spatial patterns across sessions~\cite{wang_test-retest_2022,zhang_variations_2023}, the consistency of component\slash scalp-pattern decompositions (including sign\slash scale ambiguities)~\cite{maswanganyi_statistical_2022,croce_eeg_2020,zulliger_within_2022}, and agreement of localization- or map-like outcomes when analyses are repeated across time or conditions~\cite{saha_intra-_2020,melnik_systems_2017}. In practice, spatial correlation should be interpreted alongside sign\slash scale ambiguities (e.g., ICA sign flips) and should control for referencing differences when comparing across sessions or datasets~\cite{maswanganyi_statistical_2022}. Apparent spatial-pattern differences can also be amplified by session-to-session changes in electrode placement and the underlying brain regions sampled by nominally identical sensors~\cite{scrivener_variability_2022,melnik_systems_2017}.

\textbf{Temporal pattern similarity.} For ERP waveforms and other time-series representations, similarity can be quantified via cross-correlation, waveform correlation within defined windows, or measures that explicitly accommodate latency jitter. These approaches are particularly relevant when intra-subject variability expresses itself as latency shifts (e.g., P300 latency drift) rather than simple amplitude rescaling~\cite{saha_intra-_2020,wang_test-retest_2022,ganin_sources_2023}. More generally, reliability work highlights that temporal-pattern stability is sensitive to preprocessing choices, artifact handling, scoring procedures, and state fluctuations, which can introduce structured session-to-session differences even when within-session SNR is high~\cite{wang_test-retest_2022,maswanganyi_statistical_2022,zhang_variations_2023,ribeiro_slow_2021}.

\textbf{Distance-based metrics.} Similarity can also be defined in terms of distance between multivariate feature vectors. For example, Euclidean distance between feature vectors $\mathbf{x}^{(1)}$ and $\mathbf{x}^{(2)}$ is:
\begin{equation}
d_{\mathrm{Euc}} \;=\; \sqrt{\sum_{i}\left(x^{(1)}_{i}-x^{(2)}_{i}\right)^2},
\end{equation}
where smaller distances indicate greater similarity. Mahalanobis distance generalizes this by incorporating feature covariance, which can be important when features have different scales or are correlated~\cite{maswanganyi_statistical_2022}. Distance-based approaches have been used to quantify session-to-session stability in BCI feature spaces and to characterize distributional shifts across datasets as a proxy for cross-dataset variability~\cite{wang_test-retest_2022,huang_discrepancy_2023,xu_cross-dataset_2020}. A key advantage of distance measures is that they can be applied directly to multivariate feature embeddings learned by models, enabling comparisons even when features are not easily interpretable~\cite{huang_discrepancy_2023,ma_large_2022,sartzetaki_beyond_2023}. In practice, however, distance estimates can be strongly affected by analysis\slash feature choices (``analytical variability''), motivating sensitivity analyses and transparent reporting of pipelines~\cite{allouch_effect_2023,maswanganyi_statistical_2022}.

\paragraph{Microstate transition analysis.}
In microstate research, beyond assessing stability of microstate \emph{topographies}, investigators often quantify stability of microstate \emph{dynamics}. Transition matrices summarize the probability of moving from one microstate class to another, and their stability can be assessed using correlation between transition probability matrices across sessions, entropy measures that summarize the regularity or complexity of transitions, or graph-theoretic descriptors of microstate transition networks~\cite{saha_intra-_2020,maswanganyi_statistical_2022,zanesco_within_2019}. The broader variability literature commonly reports that microstate-like spatial patterns can be relatively stable, whereas temporal dynamics and transition structure tend to be more state-dependent and therefore less reliable across sessions~\cite{saha_intra-_2020,croce_eeg_2020,liu_reliability_2020,zulliger_within_2022,zanesco_within_2019}. Because transition estimates depend on recording duration and segmentation choices, studies should report data length, number of states, and clustering\slash labeling procedures when interpreting reliability~\cite{maswanganyi_statistical_2022,zanesco_within_2019}.

\paragraph{Network-level stability in connectivity analyses.}
Connectivity-based variability can be evaluated at multiple organizational levels~\cite{saha_intra-_2020,pani_subject_2019,allouch_effect_2023,nakuci_within-subject_2023}:

\textbf{Edge-level stability.} The stability of individual connections (edges) can be assessed via test--retest correlation or ICC computed per edge across sessions~\cite{nakuci_within-subject_2023,pani_subject_2019}. Edge-level reliability is often limited by estimator variance and analysis choices (e.g., inverse solutions, connectivity metrics, number of electrodes), and is particularly sensitive to within-subject state changes across sessions~\cite{saha_intra-_2020,maswanganyi_statistical_2022,allouch_effect_2023}.

\textbf{Module-level stability.} Community structure (network modular organization) can be compared across sessions using partition-similarity views to test whether large-scale organization is preserved even when individual edges fluctuate~\cite{pani_subject_2019,nakuci_within-subject_2023}. Such analyses align with the broader observation that multivariate organization may be more stable than element-wise estimates in high-dimensional EEG representations~\cite{maswanganyi_statistical_2022,nakuci_within-subject_2023}.

\textbf{Global network metrics.} Graph-theoretic summaries such as clustering coefficient, characteristic path length, synchronizability, and centrality can be evaluated for reliability across sessions~\cite{pani_subject_2019,nakuci_within-subject_2023}. A recurring observation is that global metrics and/or appropriately pooled summaries may show better stability than individual edge weights, suggesting that network \emph{organization} can be more robust than specific pairwise connections when appropriately aggregated~\cite{saha_intra-_2020,nakuci_within-subject_2023}. This pattern is consistent with the idea that high-dimensional connectivity estimates are noisy at the edge level but can yield more stable low-dimensional summaries when appropriately pooled, while remaining sensitive to methodological degrees of freedom that should be explicitly reported~\cite{maswanganyi_statistical_2022,allouch_effect_2023}.

\begin{figure}[htbp]
\centering
\includegraphics[width=\textwidth]{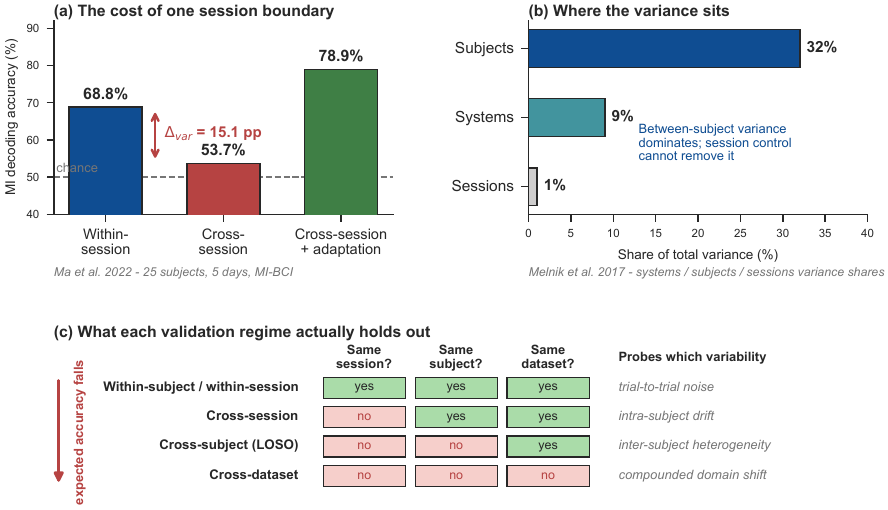}
\caption{Crossing a single session boundary costs more decoding accuracy than most methodological improvements recover, and the variance that remains is mostly between subjects. (a) In a 25-subject, 5-day motor-imagery dataset, accuracy falls from 68.8\% within-session to 53.7\% cross-session ($\Delta_{\mathrm{var}}=15.1$ percentage points, close to the 50\% chance level for two-class decoding), and online adaptation recovers it to 78.9\%~\cite{ma_large_2022}. (b) In a systems--subjects--sessions decomposition, between-subject differences account for 32\% of total variance against 9\% for recording system and 1\% for session~\cite{melnik_systems_2017}; the source does not decompose the remainder. (c) The four validation regimes differ only in what they hold out, and expected accuracy falls monotonically as more is held out, which is why a reported accuracy is uninterpretable without its regime.}
\label{fig:generalization_gap}
\end{figure}

\subsection{Machine Learning Perspectives}
Machine learning (ML) provides complementary frameworks for assessing and modeling EEG variability by treating variability as a \emph{generalization} problem: how performance and representation structure change when moving across trials, sessions, subjects, or datasets. In this view, variability is reflected in distribution shift between training and deployment conditions, and robustness is evaluated via explicit validation schemes and domain shift metrics~\cite{huang_discrepancy_2023,xu_cross-dataset_2020,kamrud_effects_2021}.

\paragraph{Cross-validation schemes as probes of variability.}
Different validation protocols isolate different sources and time scales of variability. The four schemes below differ only in what they hold out between training and testing, and expected accuracy falls monotonically as more is held out (Fig.~\ref{fig:generalization_gap}c); a reported accuracy is therefore uninterpretable without the regime that produced it.

\textbf{Within-subject cross-validation.} Training and testing on different trials or runs from the same subject (often within the same session) primarily probes trial-to-trial variability and estimation noise. This scheme typically yields the highest decoding accuracy and is most representative of single-session BCI use cases where calibration and testing occur under similar conditions~\cite{ma_large_2022,kamrud_effects_2021}.

\textbf{Cross-session validation.} Training on one session and testing on another session from the same subject probes intra-subject variability across time. Performance degradation under this scheme reflects non-stationarity, state drift, and session-level technical differences (e.g., electrode placement), and is central for evaluating long-term BCI viability~\cite{ma_large_2022,wang_test-retest_2022,zhou_relative_2021}.

\textbf{Cross-subject validation (e.g., leave-one-subject-out).} Training on a cohort of subjects and testing on a held-out subject probes inter-subject variability and is relevant to zero-calibration or low-calibration BCI aspirations. This setting typically yields the lowest accuracy because the model must generalize across large differences in spatial patterns, frequency profiles, and baseline characteristics~\cite{huang_discrepancy_2023,kamrud_effects_2021,golz_inter-individual_2019}.

\textbf{Cross-dataset validation.} Training on one dataset and testing on another evaluates a compounded form of variability that includes inter-subject heterogeneity, cross-session drift, and methodological differences across labs (hardware, protocol, preprocessing). This scheme often exposes substantial limitations in generalization and is increasingly used as a benchmark for robustness and domain adaptation methods~\cite{xu_cross-dataset_2020,sartzetaki_beyond_2023,allouch_effect_2023,ahuis_evaluation_2024}.

\paragraph{Performance degradation as an operational measure of variability.}
Many BCI studies quantify the impact of variability via performance drops between validation regimes (Fig.~\ref{fig:generalization_gap}a). A simple operational definition is:
\begin{equation}
\Delta_{\mathrm{var}} \;=\; \mathrm{Acc}_{\mathrm{within}} - \mathrm{Acc}_{\mathrm{cross}},
\end{equation}
where $\mathrm{Acc}_{\mathrm{within}}$ denotes within-subject (or within-session) accuracy and $\mathrm{Acc}_{\mathrm{cross}}$ denotes accuracy under cross-session, cross-subject, or cross-dataset testing. Larger drops indicate stronger variability effects~\cite{huang_discrepancy_2023,ma_large_2022,xu_cross-dataset_2020}. Reported comparisons often show systematic degradations when moving from within-subject to cross-session testing~\cite{ma_large_2022,zhou_relative_2021}, larger drops when moving from within-subject to cross-subject testing~\cite{huang_discrepancy_2023,kamrud_effects_2021,golz_inter-individual_2019}, and the largest reductions under cross-dataset transfer, reflecting compounded domain shift~\cite{xu_cross-dataset_2020,sartzetaki_beyond_2023}. These figures are highly dependent on dataset, preprocessing, class balance, and model choice, and should be reported with confidence intervals and consistent evaluation protocols to avoid over-comparison across studies~\cite{kamrud_effects_2021,allouch_effect_2023}.

\paragraph{Quantifying domain shift.}
Domain adaptation research provides explicit measures of distributional difference between domains (subjects, sessions, datasets), enabling variability to be quantified beyond performance outcomes~\cite{huang_discrepancy_2023,xu_cross-dataset_2020}.

\textbf{Maximum mean discrepancy (MMD).} MMD measures the distance between two distributions $P$ and $Q$ in a reproducing kernel Hilbert space:
\begin{equation}
\mathrm{MMD}(P,Q) \;=\; \left\lVert \mathbb{E}_{x\sim P}\big[\phi(x)\big] - \mathbb{E}_{y\sim Q}\big[\phi(y)\big] \right\rVert,
\end{equation}
where $\phi(\cdot)$ is a feature mapping (often implicit via a kernel)~\cite{xu_cross-dataset_2020,huang_discrepancy_2023}. In EEG, distribution-discrepancy views are used to characterize cross-session and cross-subject feature shifts and to evaluate whether adaptation reduces mismatch between training and target domains~\cite{huang_discrepancy_2023,ma_large_2022}.

\textbf{$\mathcal{A}$-distance.} The $\mathcal{A}$-distance (often used as a proxy for domain discrepancy) can be estimated from the error $\epsilon$ of a classifier trained to discriminate samples from two domains:
\begin{equation}
d_{\mathcal{A}} \;=\; 2(1 - 2\epsilon),
\end{equation}
where larger values indicate more separable domains and hence greater domain shift~\cite{kamrud_effects_2021,xu_cross-dataset_2020}. In BCI contexts, discriminability-based discrepancy views can help anticipate which subject pairs or sessions are difficult to transfer between and motivate source selection and partitioning practices that prevent optimistic generalization estimates~\cite{kamrud_effects_2021,huang_discrepancy_2023}.

Overall, ML-based perspectives treat EEG variability as a form of distribution shift that can be detected, quantified, and mitigated through careful evaluation design and explicit discrepancy measures~\cite{huang_discrepancy_2023,xu_cross-dataset_2020,sartzetaki_beyond_2023}. When combined with classical reliability statistics and variance-component models, these approaches provide a more complete methodological toolkit for understanding and managing variability in both scientific and applied EEG settings.

\subsection{Modeling Variability Explicitly}
A growing body of work argues that EEG variability should not be treated solely as nuisance noise to be removed. Instead, variability can be conceptualized as an inherent property of the neurophysiological system (state dynamics, learning, context dependence) interacting with measurement and analysis processes~\cite{saha_intra-_2020,arazi_magnitude_2016,allouch_effect_2023}. Under this view, the objective shifts from \emph{eliminating} variability to \emph{explaining, partitioning, and predicting} it.

\paragraph{Mixed-effects (multilevel) models.}
Mixed-effects models (also termed multilevel or hierarchical linear models) partition variability into fixed effects that describe population-level relationships and random effects that capture systematic subject- and session-level deviations~\cite{eckert_statistical_1974,dong_modeling_2024}. A generic formulation can be written as:
\begin{equation}
Y_{ijk} \;=\; \beta_{0} \;+\; \beta_{1}X_{ijk} \;+\; u_{0j} \;+\; u_{1j}X_{ijk} \;+\; v_{0k} \;+\; \varepsilon_{ijk},
\end{equation}

\textbf{Advantages.} Mixed-effects models are particularly well-suited to EEG variability questions because they:
\begin{itemize}
    \item explicitly represent both between-subject and within-subject (e.g., cross-session) variability via random effects~\cite{dong_modeling_2024,melnik_systems_2017};
    \item naturally handle unbalanced designs and missing data (common in EEG due to artifact-related trial loss and dropouts) without requiring listwise deletion~\cite{dong_modeling_2024,zhang_variations_2023};
    \item yield subject-specific parameter estimates (e.g., individualized slopes, learning curves), supporting individualized inference and precision modeling~\cite{dong_modeling_2024,arazi_magnitude_2016};
    \item enable hypothesis testing about moderators of variability (e.g., whether stability differs by age group, sex/gender, task, or device\slash protocol) through interactions and variance-structure comparisons~\cite{zanesco_within_2019,kumral_bold_2019,chen_cross-subject_2024}.
\end{itemize}

\textbf{Applications in EEG.} Mixed-effects and related multilevel models have been applied to quantify trial-level response variability and its group differences (e.g., evoked-response latency variability)~\cite{dong_modeling_2024,hecker_altered_2022,eyamu_prefrontal_2024}, to study how data quality and scoring choices impact ERP stability across participants and paradigms~\cite{zhang_variations_2023}, and to evaluate variability sources in multi-session or multi-factor settings where subject/session/task effects are jointly present~\cite{melnik_systems_2017,pani_subject_2019}.

\paragraph{Latent trait--state decomposition.}
Trait--state frameworks aim to disentangle stable individual characteristics from time-varying fluctuations and measurement noise:
\begin{equation}
\mathrm{Measurement}_{ij} \;=\; \mathrm{Trait}_{i} \;+\; \mathrm{State}_{ij} \;+\; \mathrm{Error}_{ij}.
\end{equation}
This decomposition is attractive for EEG because many features plausibly contain both trait-like structure and state dependence~\cite{saha_intra-_2020,zanesco_within_2019}. Evidence for trait-like stability in variability-related signatures across tasks and long time spans supports the idea that some variability magnitudes are themselves stable individual characteristics~\cite{arazi_magnitude_2016}. In parallel, microstate and spectral work indicates that whole-brain dynamics contain both within-person fluctuations and reliable between-person differences~\cite{croce_eeg_2020,liu_reliability_2020,zulliger_within_2022}.

\paragraph{Domain adaptation and domain generalization.}
Machine learning frameworks frequently conceptualize data from different subjects, sessions, or recording setups as originating from different but related \emph{domains}, each with its own distribution~\cite{huang_discrepancy_2023,xu_cross-dataset_2020}. Within this framing, variability is operationalized as \emph{domain shift}.

\textbf{Domain adaptation} assumes access to (typically unlabeled) data from the target domain and aims to align source and target distributions. In EEG, the need for such alignment is motivated by clear differences between cross-subject and cross-session feature distributions and their distinct implications for training and sample selection~\cite{huang_discrepancy_2023,ma_large_2022}. Practical strategies in the literature often combine recalibration, subject-specific transformations, or robustness-oriented preprocessing choices to reduce mismatch~\cite{arevalillo-herraez_combining_2019,ma_large_2022,allouch_effect_2023}.

\textbf{Domain generalization} aims to learn models that generalize to unseen domains without target-domain access. Multi-dataset benchmarking shows that naive pooling can overestimate performance when partitioning is imperfect, and that generalization gaps remain substantial when moving to unseen subjects or datasets~\cite{kamrud_effects_2021,xu_cross-dataset_2020,sartzetaki_beyond_2023}. Data augmentation and diversity-enhancing strategies are frequently used to improve robustness under inter-subject or inter-patient variability~\cite{aldahr_addressing_2022,goswami_improving_2024}, and methodological choices such as channel selection or common-channel identification can also target session\slash subject variability in practical pipelines~\cite{fauzi_defining_2022,changoluisa_electrode_2018}.

\paragraph{Bayesian hierarchical models.}
Bayesian hierarchical models offer a principled probabilistic framework for representing EEG variability across multiple nested levels (population $\rightarrow$ subject $\rightarrow$ session $\rightarrow$ observation) and are conceptually aligned with multilevel variance modeling common in variability research~\cite{eckert_statistical_1974,dong_modeling_2024,nakuci_within-subject_2023}. Such hierarchies mirror the structure of EEG datasets with repeated measures and can naturally express uncertainty in the presence of unequal trial counts and heterogeneous data quality~\cite{zhang_variations_2023,allouch_effect_2023}.

\textbf{Applications in EEG.} Hierarchical perspectives are particularly relevant for longitudinal and multi-session datasets and for designs that explicitly quantify within- versus between-subject reproducibility across modalities, tasks, and time~\cite{wang_test-retest_2022,nakuci_within-subject_2023,ma_large_2022}. From a variability standpoint, these models are attractive because they jointly model multiple sources of variability and quantify uncertainty, rather than providing point estimates alone~\cite{dong_modeling_2024,eckert_statistical_1974}.

\section{Variability Across Major EEG Paradigms}
\label{sec:paradigms}

\begin{figure}[htbp]
\centering
\includegraphics[width=\textwidth]{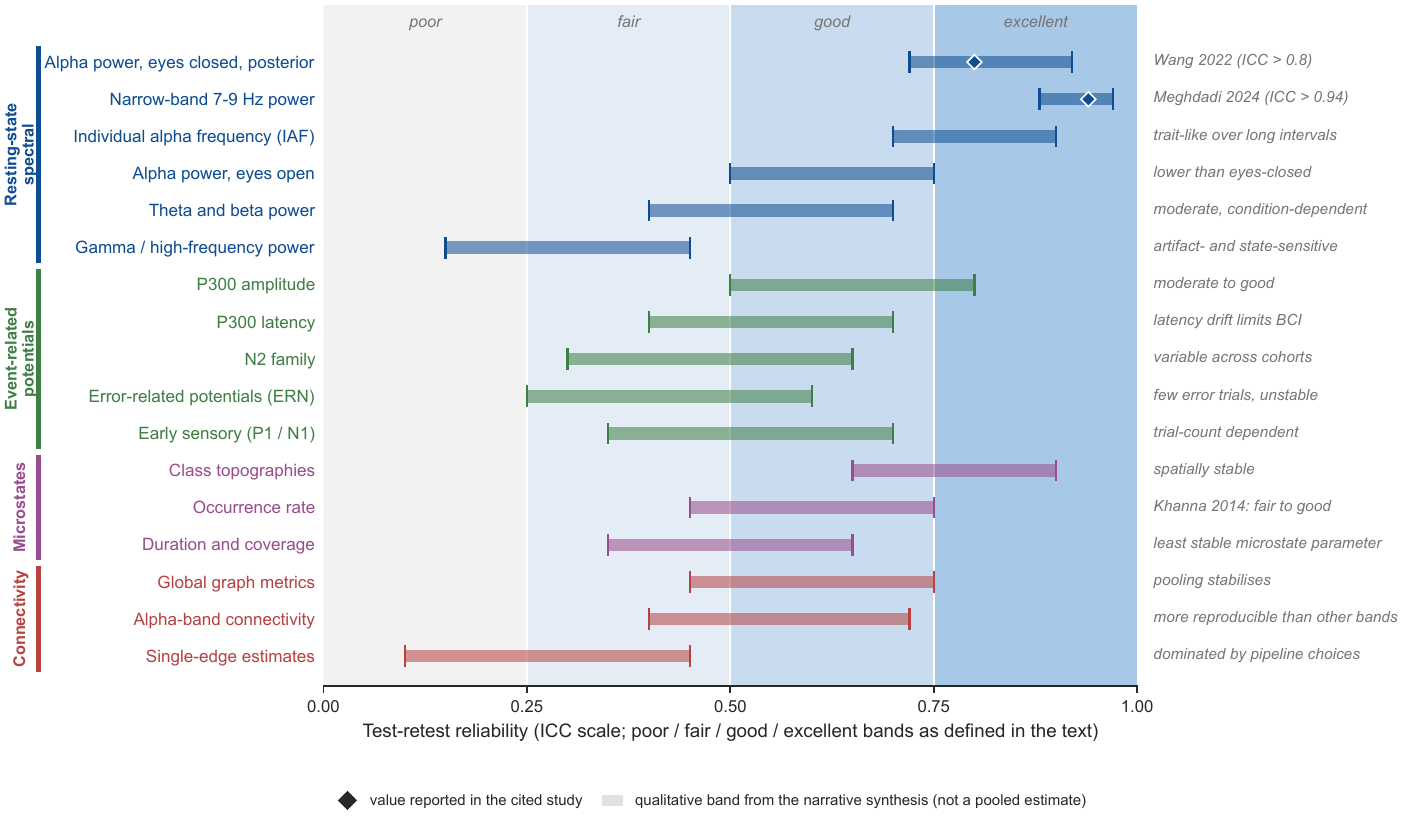}
\caption{Reliability is a property of the feature, not of EEG. Eyes-closed posterior alpha power and narrow-band 7--9~Hz power are the most stable measures in the corpus (ICC $>0.8$~\cite{wang_test-retest_2022} and ICC $>0.94$~\cite{meghdadi_inter_2024} respectively), whereas high-frequency power and single-edge connectivity estimates fall into the poor-to-fair range; within microstates, spatial topographies are consistently more stable than the temporal parameters derived from them. Diamonds mark values reported in the cited study. Bars encode the qualitative reliability band asserted for that feature in Section~\ref{sec:paradigms}, positioned on the ICC interpretation scale of Section~\ref{sec:classical_metrics}; they are an ordinal reading of the narrative synthesis and are not pooled meta-analytic estimates.}
\label{fig:reliability}
\end{figure}

This section synthesizes variability and reliability findings across major EEG paradigms (as shown in the schematic overview in Fig.~\ref{fig:reliability} and summarized study by study in Appendix~B, Table~\ref{tab:section4_variability}), with emphasis on identifying which feature families and frequency ranges tend to be more stable (and under what recording conditions) versus those that are more sensitive to state, artifacts, or methodological choices. Throughout, we distinguish (i) \emph{inter-subject variability} that supports individual-differences inference when reliable, from (ii) \emph{intra-subject variability} that limits longitudinal inference and cross-session generalization when not modeled or controlled.

\subsection{Resting-State EEG}

Resting-state EEG (rsEEG) is typically recorded during quiet wakefulness with eyes closed and/or eyes open and is widely used in cognitive neuroscience and clinical contexts due to its minimal task demands and feasibility in diverse populations~\cite{wang_test-retest_2022,nahmias_consistency_2019,melnik_systems_2017}. However, the apparent simplicity of rsEEG is accompanied by a core methodological challenge: because the paradigm imposes minimal structure, recordings are sensitive to uncontrolled mental activity and fluctuating arousal (e.g., mind wandering, drowsiness), which can introduce substantial within- and across-session variability if vigilance is not monitored or standardized~\cite{wang_test-retest_2022,ribeiro_slow_2021,parameshwaran_high_2023,zanesco_within_2019}. Consequently, rsEEG provides a useful test case for understanding how trait-like individual signatures coexist with pronounced state dependence.

\paragraph{Inter-subject variability in resting-state EEG.}
Inter-subject differences in rsEEG are often large. Large clinical and longitudinal qEEG datasets report substantial dispersion in spectral power across participants and frequency bins, reflecting both biological heterogeneity and measurement\slash processing influences~\cite{nahmias_consistency_2019,van_albada_variability_2007,meghdadi_inter_2024}. Importantly, this variability is not solely measurement noise: stable between-person differences in rhythm expression and multivariate organization can support person-level phenotyping (including ``fingerprint''-like identification) when measurements are sufficiently reliable and analysis choices are controlled~\cite{arazi_magnitude_2016,nakuci_within-subject_2023,pani_subject_2019}. Simultaneous EEG-fMRI studies show that individual alpha rhythm variations relate systematically to BOLD signals, emphasizing stable inter-subject trait differences~\cite{gonccalves2005correlating}.

\textbf{Spectral power.} Among canonical rhythms, posterior alpha activity is frequently reported to provide strong, reproducible individual differences and relatively favorable reliability compared with many other bands, motivating its use for trait-like characterization~\cite{meghdadi_inter_2024,croce_eeg_2020,arazi_magnitude_2016}. Nonetheless, alpha expression varies markedly across individuals, with some participants showing strong, well-defined posterior alpha and others displaying weaker or broader alpha activity~\cite{croce_eeg_2020,arazi_magnitude_2016}. Reported determinants of this inter-subject heterogeneity include system- and acquisition-related factors (e.g., differences in montage\slash electrode placement and the cortical regions sampled beneath electrodes)~\cite{scrivener_variability_2022,melnik_systems_2017}, variability in alpha-related spectral organization and its coupling to global brain-state structure (e.g., microstate--spectral relationships)~\cite{croce_eeg_2020,zulliger_within_2022}, recording condition (eyes open vs.\ eyes closed)~\cite{wang_test-retest_2022,parameshwaran_high_2023}, and demographic\slash clinical factors including age and cognitive status~\cite{kumral_bold_2019,ribeiro_slow_2021,eyamu_prefrontal_2024,meghdadi_inter_2024}. In contrast to alpha, theta, beta, and especially high-frequency activity (gamma) often show larger relative dispersion and stronger susceptibility to state and artifact contributions, complicating individual-differences interpretation without careful control and preprocessing~\cite{van_albada_variability_2007,golz_inter-individual_2019,allouch_effect_2023}.

\textbf{Individual alpha frequency (IAF).} IAF-related characteristics are commonly treated as relatively trait-like alpha markers, with multiple reports emphasizing substantial within-person stability over extended intervals in healthy adults and comparatively large, stable between-person differences in alpha-related dynamics~\cite{arazi_magnitude_2016,meghdadi_inter_2024,wang_test-retest_2022}. At the same time, alpha-range features can still be modulated by arousal, ongoing-signal fluctuations, and recording context, reinforcing the need to interpret IAF-style metrics within a broader state\slash measurement framework~\cite{ribeiro_slow_2021,wang_test-retest_2022}.

\textbf{Functional connectivity.} Resting-state connectivity measures (e.g., coherence, phase-synchronization indices, correlation\slash covariance-based connectivity, and graph measures) show substantial inter-subject variability and can exhibit subject-specific patterns that support person identification in multivariate network structure~\cite{pani_subject_2019,nakuci_within-subject_2023,allouch_effect_2023}. However, reliability of connectivity metrics varies strongly with the chosen estimator, frequency band, epoch length, and handling of volume conduction; analytical degrees of freedom (e.g., electrode count, source reconstruction choices, and connectivity metric selection) can substantially change both between- and within-subject similarity estimates~\cite{allouch_effect_2023,melnik_systems_2017}. Evaluation of wPLI and PLI reproducibility across multiple years confirms that connectivity metrics vary in their test-retest reliability and inter-subject variability, and that epoch length and spatial scale significantly impact stability~\cite{hardmeier2014reproducibility}. In several longitudinal network analyses, alpha-band connectivity is reported as comparatively more reproducible than connectivity derived from other frequency bands, though results remain method-dependent~\cite{nakuci_within-subject_2023}.

\paragraph{Intra-subject variability in resting-state EEG.}
Within individuals, rsEEG features vary across time scales ranging from within-session fluctuations to test--retest differences across days or months. Multiple datasets and longitudinal analyses have explicitly quantified such within-person variability across both short-term and longer-term retest intervals~\cite{wang_test-retest_2022,meghdadi_inter_2024,liu_resting_2024,arazi_magnitude_2016}.

\textbf{Test--retest reliability of spectral power.} Across studies, alpha power tends to show the most favorable test--retest profile, with good-to-excellent reliability often reported under eyes-closed conditions and at posterior electrodes in healthy cohorts~\cite{meghdadi_inter_2024,wang_test-retest_2022}. Eyes-open alpha commonly yields lower stability in many protocols, consistent with reduced alpha amplitude and increased influence of visual attention\slash fixation behavior and arousal variation~\cite{wang_test-retest_2022,parameshwaran_high_2023}. Reliability in other bands is more heterogeneous: theta and beta features can be moderately stable under controlled conditions, while very low-frequency and high-frequency ranges are often more affected by artifacts, preprocessing, and state dependence~\cite{wang_test-retest_2022,van_albada_variability_2007,golz_inter-individual_2019}.

\textbf{Stability across time scales.} Reliability can degrade as session intervals lengthen, but the relationship depends on protocol structure, familiarity\slash adaptation effects, and the degree to which session-to-session state drift is controlled or modeled~\cite{wang_test-retest_2022,meghdadi_inter_2024,arazi_magnitude_2016}. More generally, slow ongoing-signal fluctuations and arousal dynamics can confound apparent stability if not accounted for, motivating explicit adjustment or stratification by arousal-related markers~\cite{ribeiro_slow_2021}.

\textbf{State-related fluctuations during rest.} A major contributor to intra-subject variability in rsEEG is fluctuation in arousal and vigilance during ``rest,'' including transient drowsiness and mind wandering~\cite{wang_test-retest_2022,ribeiro_slow_2021,parameshwaran_high_2023}. Studies incorporating auxiliary state measures and/or explicit state modeling highlight that ongoing-signal dynamics can strongly modulate evoked and spontaneous EEG features, and that adjusting for these fluctuations can change conclusions about variability and stability~\cite{ribeiro_slow_2021,wang_test-retest_2022}. These findings motivate either (i) stricter vigilance control and standardized instructions or (ii) explicit state modeling (e.g., excluding drowsy epochs; stratifying by arousal markers) when aiming for trait-like rsEEG estimates~\cite{wang_test-retest_2022,parameshwaran_high_2023}.

\paragraph{Factors influencing resting-state variability.}
Several methodological factors recurrently modulate rsEEG variability and reliability:

\textbf{Recording duration.} Longer recordings generally yield more stable spectral and multivariate estimates by reducing estimator variance and increasing the amount of artifact-free data available for averaging and robust estimation; this logic is consistent with reliability\slash data-quality analyses emphasizing the dependence of stability on data quantity and noise levels~\cite{zhang_variations_2023,wang_test-retest_2022,meghdadi_inter_2024}.

\textbf{Eyes open versus eyes closed.} Eyes-closed rsEEG typically produces stronger alpha rhythms and often higher stability for alpha power compared to eyes-open recordings~\cite{wang_test-retest_2022,meghdadi_inter_2024}. Eyes-open conditions, however, can support more standardized visual input (e.g., fixation) and may reduce certain forms of uncontrolled imagery, though they can also increase sensitivity to attentional fluctuations depending on protocol~\cite{wang_test-retest_2022,parameshwaran_high_2023}.

\textbf{Instructions and mental context.} Instruction sets and compliance (e.g., relax vs.\ fixate, allowance of mind wandering) can alter both mean spectral profiles and within-session fluctuations. Differences in task context (rest vs.\ simple fixation) and uncontrolled cognitive content contribute to heterogeneity in reported variability ranges~\cite{wang_test-retest_2022,zanesco_within_2019,melnik_systems_2017}. Reporting exact instructions and compliance monitoring is therefore essential in variability-focused rsEEG research~\cite{melnik_systems_2017}.

\textbf{Electrode density and spatial sampling.} Higher-density montages can improve spatial characterization of topographies and connectivity patterns, potentially stabilizing multivariate estimates; however, greater preparation complexity can increase susceptibility to impedance variability and session-to-session placement differences. Empirical work shows that electrode count and placement variability, as well as downstream analysis choices, can meaningfully impact estimated similarity and reproducibility of connectivity and other multivariate features~\cite{scrivener_variability_2022,allouch_effect_2023}.

\paragraph{Clinical implications.}
Resting-state EEG is widely used within quantitative EEG (qEEG) frameworks and has been proposed as a biomarker substrate across multiple clinical contexts, supported by large clinical datasets that show both robust structure and substantial heterogeneity in qEEG features~\cite{nahmias_consistency_2019,nahmias_learning_2017}. However, substantial inter- and intra-subject variability imposes constraints on diagnostic sensitivity and longitudinal monitoring, particularly when recordings are short, vigilance is uncontrolled, or pipelines differ across sites and laboratories~\cite{meghdadi_inter_2024,liu_resting_2024,ahuis_evaluation_2024}. In clinical epilepsy studies, prolonged intracranial EEG recordings show significant variability in the rate and location of high-frequency oscillations across days, suggesting that short-duration samples can be unrepresentative~\cite{gliske2018variability}. Because normative comparisons are only as valid as the stability and harmonization of the underlying features, clinical translation requires careful attention to measurement reliability, standardized acquisition, and transparent reporting of preprocessing, referencing, and methodological choices known to induce analytical variability~\cite{allouch_effect_2023,scrivener_variability_2022,ahuis_evaluation_2024}.

\subsection{Evoked Potentials and Event-Related Potentials (ERPs)}
Event-related potentials (ERPs) are voltage deflections extracted from EEG by time-locking to discrete events (stimulus onset, response execution, feedback) and averaging across repeated trials. ERPs have been widely used to study perceptual and cognitive processes and are central to multiple clinical and BCI applications~\cite{saha_intra-_2020,changoluisa_electrode_2018,abu-alqumsan_invariance_2017,ganin_sources_2023}. Averaging improves signal-to-noise ratio by attenuating uncorrelated noise and some forms of trial-to-trial variability; however, systematic variability remains and can be substantial across individuals, across sessions, and across contexts~\cite{zhang_variations_2023,bland_exploiting_2011,ribeiro_slow_2021}. ERP variability therefore provides a useful lens for distinguishing (i) stable trait-like individual differences in cognitive processing from (ii) state-dependent fluctuations and measurement-related instability~\cite{arazi_magnitude_2016,zanesco_within_2019}.

\paragraph{Inter-subject variability in ERPs.}
Between-person variability in ERP amplitudes and latencies is widely documented and has important implications for both theory-driven cognitive neuroscience and translation-oriented biomarker research~\cite{saha_intra-_2020,zhang_variations_2023}. Inter-subject differences reflect heterogeneity in neural generators, cognitive strategy, attentional engagement, and demographic\slash clinical factors, as well as anatomical influences on scalp-recorded amplitude scaling.

\textbf{P300/P3.} The P300 elicited in oddball and related paradigms is among the most extensively studied ERP components and is foundational to several BCI pipelines~\cite{changoluisa_electrode_2018,ganin_sources_2023}. Studies report large inter-individual dispersion in P300 amplitude and latency, and emphasize that performance and interpretation can be limited by both physiological heterogeneity and component-level variability (including latency variability)~\cite{ganin_sources_2023,zhang_variations_2023}. In aging-related contexts, changes in ongoing dynamics and their modulation of evoked responses complicate between-person comparisons, reinforcing the need to control for state and arousal when attributing differences to trait or pathology~\cite{ribeiro_slow_2021,kumral_bold_2019}. In clinical\slash biomarker contexts, intra-individual ERP variability and its asymmetry (e.g., prefrontal ERP variability) has been investigated as a complementary marker beyond mean amplitude\slash latency alone~\cite{eyamu_prefrontal_2024}.

\textbf{N2.} Components in the N2/N200 family show notable inter-individual variability, and single-trial or intra-individual variability in these components has been highlighted in developmental and clinical cohorts, suggesting that dispersion may reflect meaningful neurocognitive instability rather than only measurement noise~\cite{magnuson_increased_2020,hecker_altered_2022}.

\textbf{Error-related potentials (ERN and related).} Error-related potentials can show strong between-person differences, and work on interaction error-related potentials highlights that invariance\slash variability has direct consequences for classification and generalization in error-monitoring settings~\cite{abu-alqumsan_invariance_2017}. These observations motivate reporting both average ERN-like effects and variability-aware metrics when ERPs are used for individual-level inference~\cite{maswanganyi_statistical_2022}.

\textbf{Early sensory components (P1/N1 and related).} Early visual-evoked components can exhibit substantial inter-subject variability, and recent large-scale modeling work emphasizes that trial-level variability in interpretable features (e.g., latency) can be a key differentiator in neurodevelopmental conditions~\cite{dong_modeling_2024}. More broadly, ERP data quality varies substantially across participants and paradigms, and these differences can dominate apparent between-person variability if not explicitly quantified and controlled~\cite{zhang_variations_2023}.

\paragraph{Intra-subject variability in ERPs.}
ERP intra-subject variability manifests both as trial-to-trial fluctuation within a session and as test--retest differences across sessions. Reliability evidence is heterogeneous across components and tasks, reflecting differences in SNR, trial availability, and state dependence~\cite{saha_intra-_2020,wang_test-retest_2022,zhang_variations_2023}.

\textbf{Test--retest reliability.} Test--retest work in resting and cognitive EEG shows that state changes across sessions can meaningfully alter EEG features even when within-session data quality is acceptable, motivating reliability reporting for ERP-derived features whenever longitudinal or individual-differences conclusions are drawn~\cite{wang_test-retest_2022,maswanganyi_statistical_2022}. In practice, reliability improves when sufficient artifact-free trials are available and when quantification procedures are robust to scoring choices and latency estimation issues~\cite{zhang_variations_2023}. For error-related components, variability\slash invariance properties directly affect classifier performance, and limited effective trial counts (e.g., few errors) can lead to unstable estimates that inflate apparent session-to-session differences~\cite{abu-alqumsan_invariance_2017,zhang_variations_2023}.

\textbf{Trial-to-trial variability.} Even when averaged ERPs appear stable, single-trial ERP amplitudes and latencies vary substantially around the mean waveform~\cite{arazi_magnitude_2016,bland_exploiting_2011}. Contributors include fluctuations in attention and arousal and the interaction between ongoing activity and evoked responses, which can change across time and age~\cite{ribeiro_slow_2021,parameshwaran_high_2023}. A growing body of work argues that variability is structured and behaviorally relevant: trial-by-trial neural variability magnitude can be highly reproducible across tasks and over long intervals in adults, suggesting trait-like variability signatures alongside state sensitivity~\cite{arazi_magnitude_2016}. Increased or altered intra-individual ERP variability has also been reported in clinical and developmental contexts, including autism-related cohorts and mild cognitive impairment, supporting variability-aware ERP features as complementary markers~\cite{milne_increased_2011,magnuson_increased_2020,hecker_altered_2022,eyamu_prefrontal_2024}.

\paragraph{Factors influencing ERP variability.}
Across studies, several recurring factors modulate ERP variability and reliability:

\textbf{Number of trials and SNR.} Increasing trial counts improves SNR and typically enhances stability of ERP estimates, but the realized benefit depends strongly on paradigm, participant-specific data quality, and scoring procedures~\cite{zhang_variations_2023,bland_exploiting_2011}. In BCI contexts, explicitly addressing latency variability and optimizing repetitions can improve performance, illustrating how variability-aware design choices can trade off time-on-task and estimation reliability~\cite{ganin_sources_2023}.

\textbf{Task design parameters.} Differences in cognitive demands and internal state (even within nominally similar tasks) can induce structured changes in evoked responses across sessions, complicating synthesis across studies unless task parameters and compliance are well characterized~\cite{wang_test-retest_2022,saha_intra-_2020,ribeiro_slow_2021}.

\textbf{Electrode and region-of-interest selection.} Measurement depends strongly on spatial sampling and extraction choices. Electrode selection strategies have been explicitly proposed to mitigate inter- and intra-subject variability in ERP-based BCIs, highlighting that spatial choices can materially affect robustness~\cite{changoluisa_electrode_2018}. More generally, participant- and paradigm-dependent variation in data quality implies that fixed ROI/electrode choices can differentially penalize some individuals unless justified or individualized~\cite{zhang_variations_2023}.

\textbf{Analysis decisions.} Peak vs.\ mean amplitude, time-window definitions, baseline correction, and scoring\slash latency estimation procedures can materially influence ERP estimates and their apparent stability. Systematic comparisons show that both paradigm and scoring procedure substantially affect ERP data quality, motivating sensitivity analyses and transparent reporting in variability-focused ERP work~\cite{zhang_variations_2023,maswanganyi_statistical_2022}.

\paragraph{Implications for cognitive and clinical studies.}
ERP variability has direct consequences for inference and translation.

\textbf{Statistical power and reproducibility.} High dispersion and imperfect reliability attenuate observable effects and increase sample size needs, especially when participant-level data quality and scoring procedures introduce additional variance~\cite{zhang_variations_2023,maswanganyi_statistical_2022}.

\textbf{Biomarker development.} Variability-aware ERP features (e.g., trial-level latency variability or prefrontal ERP variability measures) have been explored as potential complementary biomarkers in neurodevelopmental and cognitive-impairment contexts, but their usefulness depends on robust estimation under realistic noise and trial constraints~\cite{dong_modeling_2024,eyamu_prefrontal_2024}.

\textbf{Longitudinal inference.} Longitudinal or intervention studies require components and quantification pipelines with sufficiently high stability to distinguish true within-person change from measurement error and state-driven fluctuations. Evidence that ongoing-state fluctuations can modulate evoked responses, and that cognitive-state differences can emerge across sessions, reinforces the importance of protocol standardization and state monitoring~\cite{wang_test-retest_2022,ribeiro_slow_2021,maswanganyi_statistical_2022}.

\subsection{Oscillatory Task-Related EEG and BCI Paradigms}
Task-related oscillatory EEG activity is a central signal source for non-invasive BCI, particularly in motor imagery (MI) and sensorimotor rhythm (SMR) paradigms~\cite{saha_intra-_2020,saha_evidence_2018}. These paradigms typically exploit event-related desynchronization (ERD; task-related power decrease) and event-related synchronization (ERS; power increase, often post-movement) in mu/alpha and beta ranges over sensorimotor cortices~\cite{saha_intra-_2020,saha_evidence_2018,rimbert_is_2022}. Although oscillatory features can be robust within well-controlled sessions, MI/SMR-BCI is widely recognized as strongly impacted by both inter- and intra-subject variability, which limits generalization and affects user experience in practice~\cite{saha_intra-_2020,huang_discrepancy_2023,maswanganyi_statistical_2022}.

\paragraph{Inter-subject variability in motor imagery and BCI.}
Inter-subject variability is a defining challenge for MI-BCI and is a major reason why many systems require subject-specific calibration and have difficulty achieving ``plug-and-play'' operation~\cite{saha_intra-_2020,huang_discrepancy_2023,xu_cross-dataset_2020}. Canonical ERD/ERS responses include mu/alpha (8--13~Hz) and beta (13--30~Hz) ERD over sensorimotor cortex during imagery or movement, with beta rebound (ERS) often observed after movement cessation~\cite{saha_evidence_2018,saha_intra-_2020}. However, the magnitude, spatial distribution, temporal dynamics, and frequency specificity of these responses vary substantially across individuals~\cite{saha_intra-_2020,huang_discrepancy_2023,wriessnegger_inter-_2020}. Consequently, discriminative features can differ in optimal frequency band, electrode location, and spatial-filter subspace from one user to another, even under identical task instructions~\cite{maswanganyi_statistical_2022,zhou_relative_2021}.

Several sources contribute to inter-subject variability in MI/SMR paradigms:
\begin{itemize}
    \item \textbf{Anatomical and electrode-location factors.} Between-person differences in head\slash brain geometry and the practical variability of electrode placement relative to underlying cortex can alter the apparent amplitude and topography of sensorimotor rhythms at the scalp~\cite{scrivener_variability_2022,maswanganyi_statistical_2022}.
    \item \textbf{Functional organization.} Individuals differ in how motor imagery recruits sensorimotor networks and in the degree of lateralization during unilateral imagery, leading to strongly lateralized ERD in some users and more bilateral\slash diffuse modulation in others~\cite{saha_intra-_2020,wriessnegger_inter-_2020}.
    \item \textbf{Cognitive strategies and compliance.} Different imagery strategies (e.g., kinesthetic vs.\ visual imagery), vividness, and timing can lead to distinct oscillatory signatures and different levels of trial-to-trial consistency~\cite{saha_intra-_2020,zhou_relative_2021}.
    \item \textbf{Baseline neurophysiological traits.} Baseline rhythm strength and relative power in relevant bands\slash stages can influence the magnitude and detectability of ERD/ERS and can correlate with MI decoding performance across subjects and across time~\cite{zhou_relative_2021,rimbert_is_2022}.
\end{itemize}

A practical manifestation of strong inter-subject heterogeneity is the existence of a non-trivial subset of users who fail to achieve adequate control under standard MI protocols, reflecting a mismatch between canonical feature assumptions and the individual's true oscillatory patterns~\cite{saha_intra-_2020,huang_discrepancy_2023,borgheai_multimodal_2024}. Large-scale datasets like the OpenBMI database have been developed specifically to investigate this BCI illiteracy phenomenon across many subjects and sessions~\cite{lee2019eeg}. Importantly, recent evidence suggests that poor performance is not necessarily due to an inability to generate ERD/ERS, but can reflect how cross-subject feature distributions differ from within-subject consistency and how conventional pipelines fail to capture subject-specific discriminative structure~\cite{huang_discrepancy_2023,maswanganyi_statistical_2022}. This motivates individualized band\slash channel selection and alternative representations designed to be more robust to inter-subject differences~\cite{fauzi_defining_2022,saha_can_2023}.

Inter-subject variability is also reflected in generalization performance. Cross-subject decoding typically underperforms within-subject decoding, mirroring distributional mismatch across participants~\cite{huang_discrepancy_2023,saha_intra-_2020}. More broadly, cross-participant and cross-dataset EEG modeling can be severely biased by partitioning choices and dataset shift, leading to overestimated performance when evaluation protocols are not designed for true generalization~\cite{kamrud_effects_2021,xu_cross-dataset_2020,sartzetaki_beyond_2023}.

\paragraph{Intra-subject variability in motor imagery and BCI.}
Within a given user, MI/SMR features often drift over time, and cross-session variability is a critical limitation for practical BCI deployment~\cite{saha_intra-_2020,maswanganyi_statistical_2022,ma_large_2022}. A common observation is that a classifier trained on one session performs worse on later sessions from the same user even when tasks and equipment are nominally unchanged, consistent with non-stationarity and distribution shift in oscillatory features~\cite{huang_discrepancy_2023,ma_large_2022}. Large multi-day datasets and benchmarks further demonstrate that cross-session performance can degrade substantially, and that adaptation can recover performance, highlighting the practical importance of modeling cross-session drift~\cite{ma_large_2022}.

This drift can be decomposed into multiple contributors:
\begin{itemize}
    \item \textbf{Electrode placement and contact variability.} Session-to-session differences in cap placement and electrode contact alter spatial covariance structure and can impact spatial filtering and band-power features~\cite{scrivener_variability_2022,maswanganyi_statistical_2022}.
    \item \textbf{State changes (fatigue/drowsiness/stress).} Attention, motivation, fatigue, and arousal-related fluctuations modulate oscillatory dynamics and can induce within-person variability that is not captured by stationary models~\cite{chin-teng_lin_eeg-based_2008,golz_inter-individual_2019,hwang_mitigating_2021}.
    \item \textbf{Learning and adaptation.} Practice effects can improve MI control, but learning trajectories are heterogeneous and can interact with non-stationarity, producing non-monotonic changes across sessions~\cite{saha_intra-_2020,ma_large_2022}.
    \item \textbf{Non-stationarity and feature drift.} Even within-session EEG statistics can evolve, and these dynamics can accumulate into cross-session shifts that degrade decoders trained under stationarity assumptions~\cite{huang_discrepancy_2023,maswanganyi_statistical_2022,kamrud_effects_2021}.
\end{itemize}

In addition to cross-session effects, substantial trial-to-trial variability exists within sessions for MI-related ERD/ERS patterns~\cite{saha_evidence_2018,wriessnegger_inter-_2020}. Such variability reflects fluctuations in imagery vividness and timing and broader state dynamics, and can be behaviorally meaningful rather than purely noise~\cite{arazi_magnitude_2016,garcia_alanis_devils_2023}. Consequently, single-trial decoding is typically more sensitive to short-term variability than block-averaged analyses, reinforcing the need for robustness in real-time control~\cite{do2020estimating,saha_intra-_2020,huang_discrepancy_2023,do2021retrosplenial}. Despite this, recent advances demonstrate that real-time robotic control at the individual finger level is achievable by optimizing noninvasive BCI decoders for individual user dynamics~\cite{ding2025eegbased}.

\paragraph{Strategies to address variability in BCI.}
Given the magnitude of inter- and intra-subject variability, the MI-BCI literature has developed multiple strategies to improve robustness:

\textbf{Calibration and adaptation.} Most systems rely on an initial calibration session to fit subject-specific spatial filters and classifiers~\cite{saha_intra-_2020}. To reduce repeated recalibration, adaptive approaches update model parameters online or incrementally to track drift and compensate for within-user non-stationarity; cross-session benchmarks show that adaptation can substantially improve performance relative to naive cross-session transfer~\cite{ma_large_2022,huang_discrepancy_2023}.

\textbf{Transfer learning and evaluation rigor.} Transfer and domain adaptation aim to reduce distribution mismatch across subjects\slash sessions; however, reliable conclusions require careful dataset partitioning and cross-dataset evaluation to avoid inflated accuracy estimates~\cite{kamrud_effects_2021,xu_cross-dataset_2020,sartzetaki_beyond_2023}.

\textbf{Robust feature engineering and selection.} Stability can be improved through individualized frequency-band\slash channel selection and methods that explicitly target inter-session and inter-subject common channels or robust discriminative subspaces~\cite{fauzi_defining_2022,maswanganyi_statistical_2022}. Analytical choices can also materially affect derived metrics (especially for connectivity-style features), motivating transparency and sensitivity analyses when robustness is a central goal~\cite{allouch_effect_2023}.

\subsubsection*{Comparison with other BCI paradigms}
\textbf{SSVEP-based BCI.} Relative to MI-BCI, SSVEP often yields strong stimulus-locked responses and high inter-trial reproducibility; group-level component methods have been proposed to explicitly maximize inter-trial reproducibility and inter-subject similarity in SSVEP data~\cite{tanaka_group_2020}. Nevertheless, SSVEP performance can still vary with user state (e.g., stress\slash fatigue) and other context factors that influence attention and visual processing~\cite{zhang_stress-induced_2020}.

\textbf{Hybrid BCIs.} Hybrid systems may improve robustness by combining complementary control signals, but can also inherit variability sources from each component paradigm and increase calibration\slash workload complexity~\cite{saha_intra-_2020,maswanganyi_statistical_2022}.

\subsection{Advanced Paradigms}
Beyond conventional resting-state, ERP, and canonical oscillatory BCI paradigms, several advanced experimental and analytical approaches have been developed to better characterize EEG dynamics under realistic conditions and to explicitly probe sources of inter- and intra-subject variability~\cite{melnik_systems_2017,garcia_alanis_devils_2023}. These approaches are particularly valuable because they (i) increase ecological validity by sampling broader contexts (multiple tasks, multiple days, multiple sites), and (ii) provide richer structure for decomposing variance into trait-like and state-like components~\cite{wang_test-retest_2022,huang_m3_2022}.

\paragraph{Multi-task and multi-session datasets.}
A notable recent trend is the development of large-scale datasets that deliberately incorporate multiple tasks and repeated sessions per participant, enabling systematic quantification of within-person drift and between-person heterogeneity under a unified protocol~\cite{huang_m3_2022,wang_test-retest_2022}. Such resources support design-aware reliability estimation (e.g., how stability changes with session spacing or trial count) and facilitate benchmarking of generalization methods (transfer learning, domain adaptation, meta-learning) under standardized evaluation conditions~\cite{ma_large_2022,kamrud_effects_2021,sartzetaki_beyond_2023}.

For example, multi-session datasets in MI and related BCI settings demonstrate that (i) cross-session variability is a robust phenomenon, (ii) the magnitude and structure of variability can be paradigm- and feature-dependent, and (iii) individuals differ substantially in their learning and stability trajectories over repeated sessions~\cite{ma_large_2022,maswanganyi_statistical_2022,huang_discrepancy_2023}. More broadly, large-cohort datasets with dozens to hundreds of participants enable population-level characterization of performance variability and the study of correlates of BCI aptitude and stability, including user-profile information and state-related factors~\cite{dreyer_pauline_large_2023,saha_can_2023,borgheai_multimodal_2024}. Importantly, these datasets shift the focus from demonstrating performance within carefully curated sessions to quantifying robustness across realistic deployment conditions, where session-to-session drift and user-state fluctuations are unavoidable~\cite{melnik_systems_2017,wang_test-retest_2022}.

\paragraph{Cross-dataset generalization and ``dataset shift''.}
Cross-dataset generalization studies (training on one dataset and evaluating on another) provide a stringent test of whether models capture invariant neurophysiological structure or instead overfit dataset-specific characteristics~\cite{xu_cross-dataset_2020,sartzetaki_beyond_2023,kamrud_effects_2021}. Even for nominally similar paradigms, cross-dataset performance can degrade substantially, reflecting compounded sources of variability including (i) hardware and montage differences, (ii) protocol and instruction differences, (iii) population differences, and (iv) preprocessing and feature-extraction choices~\cite{xu_cross-dataset_2020,kamrud_effects_2021,allouch_effect_2023}. From a variability perspective, cross-dataset studies emphasize that ``robustness'' is not solely an intra-subject non-stationarity problem; it is also a domain shift problem that spans acquisition systems and analysis pipelines~\cite{melnik_systems_2017,xu_cross-dataset_2020,allouch_effect_2023}. Consequently, variability-aware benchmarking benefits from explicit reporting of dataset-level factors, careful partitioning\slash evaluation design, and sensitivity analyses that quantify how much variance is attributable to key pipeline decisions~\cite{kamrud_effects_2021,allouch_effect_2023}.

\paragraph{Microstate analysis.}
Microstate analysis characterizes EEG as a sequence of quasi-stable global scalp topographies that persist for tens to hundreds of milliseconds, providing a compact description of large-scale spatiotemporal dynamics~\cite{zanesco_within_2019,liu_reliability_2020}. Microstates are increasingly used in both basic and clinical EEG research because they summarize complex multichannel activity into interpretable temporal parameters and can be related to spectral and network-level properties~\cite{croce_eeg_2020,zulliger_within_2022}.

\textbf{Inter-subject variability.} While canonical microstate classes can show broadly consistent topographies across individuals, microstate temporal parameters frequently vary across participants. Reported sources of inter-subject variability include differences in average microstate duration, coverage (fraction of time occupied by each class), and the temporal dynamics of occurrence and sequencing~\cite{zanesco_within_2019,liu_reliability_2020,zulliger_within_2022}. Such variability may reflect stable differences in large-scale dynamics, but it is also influenced by recording context and methodological decisions~\cite{laganaro_inter-study_2017,liu_reliability_2020}.

\textbf{Intra-subject variability and reliability.} Across-session assessments generally suggest that microstate \emph{topographies} can be relatively stable, whereas \emph{temporal parameters} (e.g., duration, coverage, occurrence rate) show more moderate reliability and can vary with state and recording conditions~\cite{liu_reliability_2020,zanesco_within_2019}. Resting-state microstate features demonstrate fair to good test-retest reliability across multiple sessions, though metrics like frequency of occurrence are generally more stable than microstate duration~\cite{khanna2014reliability}. Microstate parameters are also systematically related to band-limited spectral features (especially alpha), highlighting that fluctuations in oscillatory state can covary with microstate dynamics within and across individuals~\cite{croce_eeg_2020,zulliger_within_2022}. These findings motivate detailed reporting of microstate pipeline decisions (e.g., number of states, clustering strategy, template use) and, when possible, sensitivity analyses, because methodological choices can meaningfully affect inferred stability~\cite{liu_reliability_2020,laganaro_inter-study_2017}.

\textbf{Clinical relevance.} Microstate features have been explored as candidate markers in clinical EEG research; however, moderate reliability of some temporal parameters and large inter-subject dispersion present challenges for individual-level inference, which is why standardized protocols, robust estimation, and validation in independent cohorts~\cite{liu_reliability_2020,laganaro_inter-study_2017}.

\section{Discussion}
\label{sec:discussion}
This review highlights that inter- and intra-subject EEG variability is not a single phenomenon but the observable outcome of interacting biological, psychological\slash state, technical, and analytical factors. Across paradigms, variability limits reproducibility and cross-domain generalization, but it also contains stable, person-specific structure that can be used for individual-differences neuroscience and precision neurotechnology~\cite{saha_intra-_2020,huang_discrepancy_2023,arazi_magnitude_2016,liu_reliability_2020}. A key implication is that variability should be treated as a \emph{design variable}: what a study can reliably infer depends on how recordings are repeated, how data are partitioned for evaluation, and how analytic choices are controlled and documented~\cite{kamrud_effects_2021,allouch_effect_2023}.

\subsection{Interpretation of variability across paradigms and features}
The literature consistently indicates that cross-session and cross-subject shifts are major sources of performance degradation for EEG decoding, particularly in motor imagery BCIs, where distribution changes across days and across users remain a central barrier to robust deployment~\cite{huang_discrepancy_2023,ma_large_2022,maswanganyi_statistical_2022}. At the same time, trait-like components are repeatedly observed: trial-by-trial neural variability magnitudes can be stable over long intervals and across tasks~\cite{arazi_magnitude_2016}, and microstate parameters and related spectral expressions exhibit reproducible within- and between-subject structure that supports the idea of individual ``fingerprints'' alongside state dependence~\cite{croce_eeg_2020,liu_reliability_2020,zulliger_within_2022}. These findings reinforce a dual perspective: variability is problematic when the goal is estimating small group effects under heterogeneous states, but it is informative when the goal is characterizing stable individual organization or state dynamics.

\subsection{Methodological contributors and the need for transparency}
A recurring theme is that a non-trivial fraction of observed variability is methodological. Electrode placement can vary meaningfully even with fixed caps, altering the cortical regions sampled and thereby contributing to between-subject and between-session differences~\cite{scrivener_variability_2022}. Likewise, results from EEG source connectivity and network analyses can change substantially with pipeline decisions (e.g., sensor density, inverse solution, connectivity metric), producing analytical variability that directly impacts between- and within-subject similarity and group-level consistency~\cite{allouch_effect_2023}. For cross-participant deep learning models, improper dataset partitioning can lead to severe overestimation of accuracy, so that evaluation protocols are inseparable from claims about generalization under variability~\cite{kamrud_effects_2021,xu_cross-dataset_2020}. These observations motivate a stronger norm: studies should treat analysis pipelines and validation strategies as first-class methodological objects to be justified, documented, and sensitivity-tested.

\subsection{Recommendations for study design, reporting, and analysis}
To support cumulative science and translation, EEG variability studies should treat design, reporting, and analysis as an integrated system rather than independent choices.

\paragraph{Repeated-measures designs.}
Separating trait-like subject effects from state fluctuations and measurement noise requires repeated recordings. Public resources illustrate how multi-session designs enable quantification of both short- and longer-term stability across resting and cognitive states~\cite{wang_test-retest_2022}. In BCI contexts, multi-day datasets show that cross-session classification can degrade markedly relative to within-session evaluation, but that adaptation strategies and training over session diversity can improve robustness, making repeated measures essential for realistic benchmarking~\cite{ma_large_2022,huang_discrepancy_2023}. When feasible, multi-task and multi-session collections can further characterize how task-switching compares to session effects and subject-specific structure~\cite{huang_m3_2022,pani_subject_2019}.

\paragraph{Standardization and metadata for controllable sources of variance.}
Avoidable variability should be reduced through procedural standardization and explicit metadata capture. In addition to conventional reporting of montage and preprocessing, studies should record electrode placement/fit information when possible, given demonstrated placement variability and its potential impact on sampled brain regions~\cite{scrivener_variability_2022}. For multi-laboratory or multi-site settings, evidence from coordinated preclinical work shows that harmonization of protocols and centralized analysis can reduce between-laboratory variance, supporting investment in standardized procedures, training, and centralized QC~\cite{ahuis_evaluation_2024}. Rich subject- and state-level profiling is also valuable for interpreting performance variation and enabling covariate modeling; datasets with behavioral\slash psychological measures and subject-driven states provide a template for this approach~\cite{wang_test-retest_2022}.

\paragraph{Reliability reporting and uncertainty.}
Reliability should be reported whenever EEG measures are used for longitudinal inference, individual differences, biomarker development, or patient-level interpretation. Large clinical datasets demonstrate that quantitative EEG feature consistency can be characterized at scale, providing benchmarks for what is stable in realistic settings~\cite{nahmias_consistency_2019}. For longitudinal effect interpretation, recent work emphasizes that inter- and intra-subject variability can meaningfully alter normalized effect-size estimates across frequencies, implying that both population variability and change-score variability should be considered when interpreting intervention effects~\cite{meghdadi_inter_2024,liu_resting_2024}. More broadly, component- and procedure-dependent ERP data quality varies strongly across paradigms and participants, implying that reliability and measurement error should be routinely quantified and discussed rather than assumed~\cite{zhang_variations_2023}.

\paragraph{Evaluation under realistic domain shifts.}
Given the prevalence of cross-session and cross-dataset degradation, claims about generalization should be anchored in evaluations that explicitly test shifts across sessions, subjects, and datasets. Cross-dataset variability is a documented failure mode for deep learning decoding, and cross-participant evaluation can be misleading without strict separation of subjects across data splits~\cite{xu_cross-dataset_2020,kamrud_effects_2021}. Therefore, studies proposing ``general'' models should report leave-one-subject-out or equivalent subject-independent testing, and where possible include cross-session and cross-dataset transfer as primary (not secondary) outcomes~\cite{golz_inter-individual_2019,sartzetaki_beyond_2023}.

\begin{figure}[htbp]
\centering
\includegraphics[width=\textwidth]{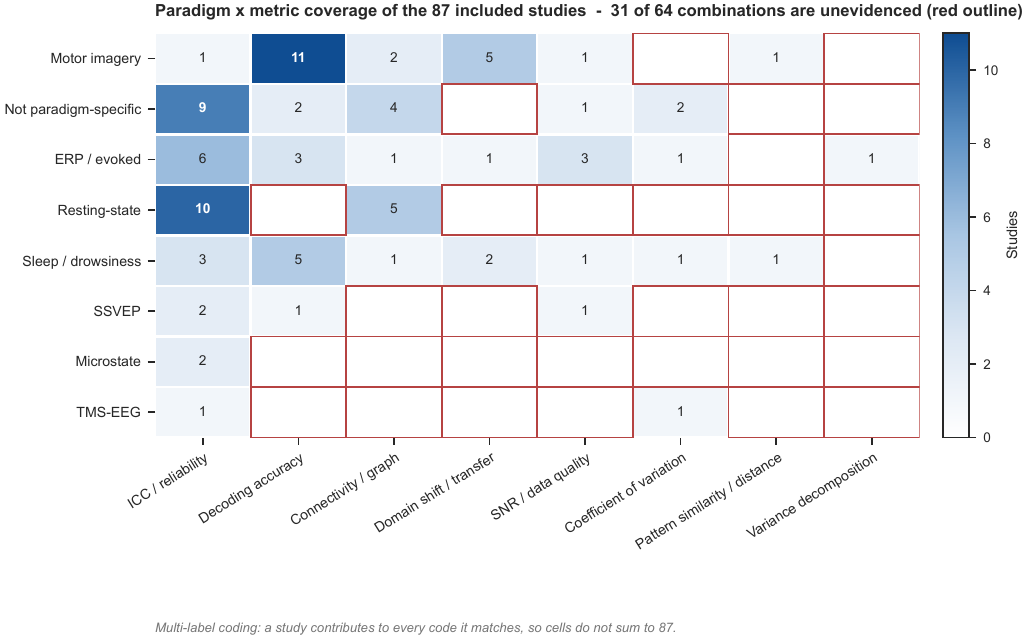}
\caption{Paradigm and metric are confounded across the corpus, and the gaps are systematic rather than incidental. Motor-imagery studies overwhelmingly report decoding accuracy while resting-state studies report ICC, so the two literatures rarely produce comparable numbers for the same feature. Thirty-one of the 64 paradigm $\times$ metric combinations carry no study at all (red outline), including every combination of microstate or SSVEP paradigms with variance decomposition. Coding is multi-label, so cells do not sum to 87; ``not paradigm-specific'' collects the cross-paradigm and purely methodological studies.}
\label{fig:evidence_map}
\end{figure}

\subsection{Future directions}
The coverage map in Fig.~\ref{fig:evidence_map} makes the field's structural gaps explicit and organizes the directions below: the corpus is thick where a paradigm has a native performance metric and thin wherever a claim requires comparing across paradigms, sites, or datasets.

\paragraph{Large, repeated, and diverse datasets.}
A central constraint in the variability literature is the limited availability of large, multi-session datasets with rich phenotyping. Emerging multi-subject\slash multi-session resources for MI and broader multi-task databases provide important building blocks for robust variance decomposition and benchmarking~\cite{ma_large_2022,huang_m3_2022}. Extending this direction to multi-site collections, and pairing them with harmonized acquisition and centralized QC, is likely to be decisive for quantifying site effects and enabling reproducible cross-domain generalization~\cite{ahuis_evaluation_2024}. Such datasets should also include detailed device\slash protocol metadata and, when feasible, measures that index subject state (sleep, fatigue, stress) given evidence that these factors relate to performance variation~\cite{zhou_relative_2021,zhang_stress-induced_2020,borgheai_multimodal_2024}.

\paragraph{From paradigm-specific engineering to representation learning.}
The variability challenge can be reframed as a representation problem: models should learn features that preserve neurophysiological invariants while discounting nuisance variation. Evidence of predictable performance variability (e.g., links with spectral state markers) suggests that part of the ``noise'' may be modeled explicitly or used for adaptive control policies~\cite{zhou_relative_2021,borgheai_multimodal_2024}. In addition, multi-dataset fine-tuning studies indicate both the promise and pitfalls of transfer, motivating benchmarks that quantify gains under realistic shifts rather than within-dataset tuning alone~\cite{sartzetaki_beyond_2023}. Progress here will depend on transparent cross-session and cross-dataset evaluations and on reporting practices that prevent inadvertent leakage or overestimation~\cite{kamrud_effects_2021}.

\paragraph{Edge computing for more reproducible EEG pipelines.}
A practical future direction to reduce system-induced variability is to move parts of the EEG processing and inference pipeline from heterogeneous host computers to standardized embedded\slash edge platforms. By keeping acquisition interfacing, preprocessing (e.g., filtering\slash re-referencing), feature extraction, and decoding within a fixed hardware\slash software stack, edge deployment can improve run-to-run determinism (timing, latency, and numerical consistency) and reduce variability introduced by differences in operating systems, drivers, compute loads, and network connectivity. While edge computing cannot remove physiological sources of variability (state fluctuations, learning, fatigue), it may meaningfully reduce avoidable engineering variability and thereby support more reliable longitudinal measurements and real-world BCI use. Recent overviews of Edge AI for BCIs and fully embedded SSVEP platforms highlight the feasibility and emerging design patterns for on-device processing in low-power real-time settings~\cite{nguyen2025edge,nguyen2026edgessvep}.

\paragraph{Variability as a clinical and cognitive signal.}
Finally, variability itself may serve as a target phenotype. Work on autism and cognitive impairment illustrates that intra-individual ERP variability can differ systematically between groups and may provide biomarker-relevant information beyond mean amplitudes\slash latencies~\cite{magnuson_increased_2020,eyamu_prefrontal_2024,dong_modeling_2024}. Similarly, dynamic variability metrics (e.g., high-variability periods; microstate dynamics) can distinguish cognitive brain states and link to individual differences, encouraging a shift from viewing variability only as error toward treating it as a mechanistic object of study~\cite{parameshwaran_high_2023,zanesco_within_2019}.

\subsection{Limitations of the present review}
\label{sec:limitations}
Several limitations follow from the review design and should be weighed when reading the synthesis. First, and by design, we did not appraise methodological quality or risk of bias, so a finding reported by a four-participant study and one reported by a hundred-participant multi-session study carry equal weight in the synthesis; readers should not treat cell counts in Fig.~\ref{fig:evidence_map} as evidence strength. Second, we did not pool estimates, and the reliability bands in Fig.~\ref{fig:reliability} are an ordinal reading of the reviewed narratives rather than meta-analytic intervals; a future review that extracts per-study ICC values with confidence intervals could replace them with a genuine forest plot. Third, the review was not prospectively registered and no protocol was published, so there is no external record fixing the review questions before the literature was read. Fourth, screening and data extraction were performed without an independently duplicated second pass or an inter-rater agreement statistic, so the coding in Figs.~\ref{fig:corpus} and~\ref{fig:evidence_map} should be read as a structured mapping rather than as an independently verified classification. 

Three further limitations concern the evidence itself rather than the review design. Nine of the 87 reviewed records are preprints, posted content, a dissertation, or a dataset descriptor and have not undergone peer review; we retained them because several are the primary description of multi-session resources on which this literature depends, but their claims carry correspondingly less weight. The included literature also spans heterogeneous paradigms, devices, preprocessing pipelines, and variability metrics, which complicates direct comparison and may itself reflect analytical variability in the field; variability estimates depend on design choices such as retest interval and task structure, so conclusions about ``stability'' are conditional on those contexts. Finally, the search was restricted to English-language records indexed in six databases, so work published elsewhere, and grey literature more broadly, is under-represented.

\section{Conclusion}
\label{sec:conclusion}
Reviewing 87 studies of inter- and intra-subject EEG variability yields one organising message: variability is pervasive, structured, and consequential, but it is a property of specific features and specific protocols rather than of EEG as a modality. Eyes-closed posterior alpha and individual alpha frequency support individual-level inference; high-frequency power, single-edge connectivity, and microstate temporal parameters do not, and a substantial share of what is reported as neural variability originates instead in electrode placement, referencing, pipeline choices, and evaluation design. The practical consequences differ by field: attenuated effect sizes and failed replications in cognitive neuroscience, calibration burden and cross-session drift in BCI, constrained individual-level decisions in clinical use. The remedy is common to all three, namely to treat variability as a design variable rather than as residual noise, report reliability and variance components as routinely as means, and evaluate generalisation under the shift that deployment will actually impose. The review also shows where this literature is not yet able to support such claims. Paradigm and metric remain confounded, so motor-imagery and resting-state findings rarely produce comparable numbers for the same feature; 31 of 64 paradigm~$\times$~metric combinations carry no study at all; and only two of the reviewed studies evaluate cross-dataset or multi-site designs, which is precisely the evidence that robustness claims require. Closing that gap will depend less on new metrics than on harmonised acquisition, transparent reporting of analytical degrees of freedom, and large multi-session and multi-site datasets that make realistic domain shift measurable rather than assumed.

\section*{Declarations}

\paragraph{Funding.}
This work was supported by Vietnam Maritime University.

\paragraph{Competing interests.}
The authors have no competing interests to declare that are relevant to the
content of this article.

\paragraph{Ethics approval.}
Not applicable. This article is a review of previously published studies and
does not report any new study with human participants or animals performed by
any of the authors.

\paragraph{Consent to participate.}
Not applicable.

\paragraph{Consent for publication.}
Not applicable.

\paragraph{Data availability.}
The evidence table for all 87 included studies is provided in Appendix~A. The
search records and the extracted study characteristics underlying Figures~2
and~7 are available from the corresponding author on reasonable request. No new
empirical data were generated for this review.

\paragraph{Code availability.}
The analysis scripts that produced the figures and the appendix evidence table
are available from the corresponding author on reasonable request.

\paragraph{Author contributions.}
X.-T.T. conceived the review, designed the search strategy, screened and
selected the included studies, extracted and coded the study records, prepared
all figures and tables, and wrote the main manuscript text. T.-N.V. and S.-T.V.
contributed to record screening and data extraction and verified the evidence
table in Appendix~A. T.-T.T. and M.-D.N. contributed to the literature search
and to the synthesis of variability metrics and paradigms. T.D. contributed to
the interpretation of the brain--computer interface and machine-learning
sections and revised those parts of the manuscript. C.-T.L. supervised the work
and critically revised the manuscript for intellectual content. All authors
reviewed and approved the final manuscript.

\section*{Appendix A: Evidence table for the 87 included studies}
\label{app:evidence}
\addcontentsline{toc}{section}{Appendix A}

\noindent
Every study included in this review is listed below in chronological order.
\textbf{Paradigm:} Rest = resting-state, ERP = event-related/evoked potential,
MI = motor imagery, SSVEP = steady-state visual evoked potential,
MS = microstate, Sleep = sleep/drowsiness/vigilance, TMS = TMS--EEG.
\textbf{Target:} Inter = inter-subject, Intra = intra-subject, Sess = cross-session,
Trial = trial-to-trial, Data = cross-dataset or multi-site.
\textbf{Metric:} ICC = intraclass correlation or test--retest reliability,
CV = coefficient of variation, SNR = signal-to-noise or data quality,
VarD = variance decomposition, Patt = pattern similarity or distance,
Acc = decoding accuracy, Shift = domain shift or transfer, Conn = connectivity or graph.
A dash means no code in that family was matched. Coding is multi-label, so a study
may carry several codes in one column and the column totals exceed 87.

\vspace{2mm}
\noindent\emph{Note.} These codes were assigned by word-boundary keyword matching over
each record's title, abstract and extracted summary fields. They provide a structured
map of the corpus, not an independently verified expert classification, and were not
used to appraise study quality (see Section~\ref{sec:limitations}).

\vspace{3mm}
{\footnotesize
\setlength{\tabcolsep}{4pt}
\renewcommand{\arraystretch}{1.12}
\begin{longtable}{r p{3.4cm} r c p{2.1cm} p{2.4cm} p{3.4cm}}
\caption{Structured record for each of the 87 studies included in the review.}
\label{tab:evidence}\\
\toprule
\textbf{\#} & \textbf{Study} & \textbf{Year} & \textbf{Ref.} & \textbf{Paradigm} & \textbf{Target} & \textbf{Metric} \\
\midrule
\endfirsthead
\toprule
\textbf{\#} & \textbf{Study} & \textbf{Year} & \textbf{Ref.} & \textbf{Paradigm} & \textbf{Target} & \textbf{Metric} \\
\midrule
\endhead
\midrule
\multicolumn{7}{r}{\emph{continued on next page}} \\
\endfoot
\bottomrule
\endlastfoot
1 & Iwahara \& Noguchi & 1972 & \cite{iwahara_intra_1972} & Rest & Inter, Intra, Sess & -- \\
2 & Eckert & 1974 & \cite{eckert_statistical_1974} & -- & Intra & ICC \\
3 & Vogel & 1986 & \cite{f_principles_1986} & ERP & Intra & -- \\
4 & Lauter \& Loomis & 1986 & \cite{lauter_individual_1986} & -- & Inter, Intra & CV \\
5 & Kohrman \textit{et al.} & 1989 & \cite{kohrman_inter-_1989} & Rest & Inter, Intra & ICC \\
6 & Ma \textit{et al.} & 1991 & \cite{ma_interoperator_1991} & -- & Intra & ICC, Conn \\
7 & Nicolaou \& Nasuto & 2005 & \cite{nicolaou_robustness_2005} & MI & Inter, Intra & SNR, Acc \\
8 & van Albada \textit{et al.} & 2007 & \cite{van_albada_variability_2007} & Rest & Trial & ICC \\
9 & Lin \textit{et al.} & 2008 & \cite{chin-teng_lin_eeg-based_2008} & Sleep & Intra, Sess & Patt \\
10 & Gonçalves \textit{et al.} & 2008 & \cite{goncalves_data_2008} & Rest & Inter, Intra & ICC, Conn \\
11 & Gonçalves \textit{et al.} & 2008 & \cite{goncalves_inter-subject_2008} & Rest & Inter, Intra & ICC \\
12 & Roijendijk & 2009 & \cite{roijendijk_variability_2009} & Sleep & Intra & -- \\
13 & Christoforou \textit{et al.} & 2010 & \cite{christoforou_bilinear_2010} & -- & Inter, Intra, Sess & -- \\
14 & Thomas \textit{et al.} & 2010 & \cite{thomas_study_2010} & MI & Sess & Acc \\
15 & Bland \textit{et al.} & 2011 & \cite{bland_exploiting_2011} & ERP & Intra, Trial & -- \\
16 & Milne & 2011 & \cite{milne_increased_2011} & -- & Inter & ICC, Conn \\
17 & Choudhury \textit{et al.} & 2011 & \cite{roy_choudhury_intra_2011} & ERP, TMS & Inter, Intra & ICC, CV \\
18 & Hossain \textit{et al.} & 2013 & \cite{hossain_canonical_2013} & ERP, SSVEP & Inter, Intra, Trial & -- \\
19 & Hitziger & 2015 & \cite{hitziger_modeling_2015} & -- & Intra & -- \\
20 & Iscan \textit{et al.} & 2016 & \cite{iscan_pre-stimulus_2016} & ERP, TMS & Inter, Intra & -- \\
21 & Abu-Alqumsan \textit{et al.} & 2017 & \cite{abu-alqumsan_invariance_2017} & -- & Inter, Intra & Acc \\
22 & Arazi \textit{et al.} & 2017 & \cite{arazi_magnitude_2016} & -- & Inter, Intra & CV \\
23 & Bodmer \textit{et al.} & 2017 & \cite{bodmer_neurophysiological_2017} & -- & Inter, Intra & -- \\
24 & Laganaro & 2017 & \cite{laganaro_inter-study_2017} & ERP, MS & Inter & ICC \\
25 & Melnik \textit{et al.} & 2017 & \cite{melnik_systems_2017} & ERP, SSVEP & Inter, Intra & ICC \\
26 & Nahmias \textit{et al.} & 2017 & \cite{nahmias_learning_2017} & -- & Inter, Intra & ICC \\
27 & Szostakiwskyj \textit{et al.} & 2017 & \cite{szostakiwskyj_modulation_2017} & Rest & Inter, Intra & -- \\
28 & Changoluisa \textit{et al.} & 2018 & \cite{changoluisa_electrode_2018} & ERP & Intra & -- \\
29 & Saha \textit{et al.} & 2018 & \cite{saha_evidence_2018} & MI & Inter, Sess & Acc, Conn \\
30 & Arevalillo-Herráez \textit{et al.} & 2019 & \cite{arevalillo-herraez_combining_2019} & -- & Inter, Intra & ICC \\
31 & Golz \textit{et al.} & 2019 & \cite{golz_inter-individual_2019} & Sleep & Inter, Intra & CV, Acc \\
32 & Kumral \textit{et al.} & 2019 & \cite{kumral_bold_2019} & -- & Intra & -- \\
33 & Murray \textit{et al.} & 2019 & \cite{murray_contributions_2019} & -- & -- & -- \\
34 & Nahmias \textit{et al.} & 2019 & \cite{nahmias_consistency_2019} & -- & Inter, Intra & ICC \\
35 & Pani \textit{et al.} & 2019 & \cite{pani_subject_2019} & -- & Inter, Intra, Sess & Conn \\
36 & Salazar-Varas \& Vázquez & 2019 & \cite{salazar-varas_facing_2019} & -- & Inter, Intra & ICC, Acc \\
37 & Zanesco \textit{et al.} & 2019 & \cite{zanesco_within_2019} & Rest, MS & Inter, Intra & -- \\
38 & Chikara \& Ko & 2020 & \cite{chikara_prediction_2020} & ERP & Inter, Intra & Acc \\
39 & Croce \textit{et al.} & 2020 & \cite{croce_eeg_2020} & Rest, MS & Inter, Intra & -- \\
40 & Liu \textit{et al.} & 2020 & \cite{liu_reliability_2020} & Rest, MS & Inter, Intra & ICC \\
41 & Magnuson \textit{et al.} & 2020 & \cite{magnuson_increased_2020} & ERP & Inter, Intra & -- \\
42 & Saha \& Baumert & 2020 & \cite{saha_intra-_2020} & MI & Inter, Intra, Sess & Shift \\
43 & Kumawat & 2020 & \cite{senior_professor_department_of_physiology_s_m_s_medical_college_jaipur_rajasthan_inter_2020} & Rest, ERP & Intra & Conn \\
44 & Tanaka & 2020 & \cite{tanaka_group_2020} & ERP, SSVEP & Inter & ICC \\
45 & Wang \textit{et al.} & 2020 & \cite{wang_spatial_2020} & Rest & Intra & ICC \\
46 & Wriessnegger \textit{et al.} & 2020 & \cite{wriessnegger_inter-_2020} & MI & Inter, Intra & Patt \\
47 & Xu \textit{et al.} & 2020 & \cite{xu_cross-dataset_2020} & MI & Inter, Intra, Data & -- \\
48 & Zhang \textit{et al.} & 2020 & \cite{zhang_stress-induced_2020} & ERP, SSVEP, Sleep & Intra & SNR, Acc \\
49 & Hwang \textit{et al.} & 2021 & \cite{hwang_mitigating_2021} & Sleep & Inter & Acc \\
50 & Kamrud \textit{et al.} & 2021 & \cite{kamrud_effects_2021} & -- & Intra & -- \\
51 & Li \textit{et al.} & 2021 & \cite{li_brain_2021} & Rest, ERP & Inter, Intra & -- \\
52 & Ribeiro \& Castelo-Branco & 2021 & \cite{ribeiro_slow_2021} & -- & Intra & -- \\
53 & Zhao \textit{et al.} & 2021 & \cite{zhao_intra-subject_2021} & -- & Inter, Intra & -- \\
54 & Zhou \textit{et al.} & 2021 & \cite{zhou_relative_2021} & MI & Inter, Intra & Acc \\
55 & Aldahr \textit{et al.} & 2022 & \cite{aldahr_addressing_2022} & -- & Intra & -- \\
56 & Fauzi \textit{et al.} & 2022 & \cite{fauzi_defining_2022} & MI & Inter, Intra, Sess & Acc \\
57 & Hecker \textit{et al.} & 2022 & \cite{hecker_altered_2022} & -- & Intra & ICC \\
58 & Huang \textit{et al.} & 2022 & \cite{huang_m3_2022} & Rest & Inter, Intra, Sess & -- \\
59 & Ma \textit{et al.} & 2022 & \cite{ma_large_2022} & MI & Inter, Intra, Sess & Acc \\
60 & Maswanganyi \textit{et al.} & 2022 & \cite{maswanganyi_statistical_2022} & ERP, MI, Sleep & Inter, Sess & ICC, Acc, Shift \\
61 & Rimbert \textit{et al.} & 2022 & \cite{rimbert_is_2022} & Rest, MI & Inter, Intra & -- \\
62 & Scrivener \& Reader & 2022 & \cite{scrivener_variability_2022} & -- & Inter, Intra & -- \\
63 & Wang \textit{et al.} & 2022 & \cite{wang_test-retest_2022} & Rest, Sleep & Sess & ICC \\
64 & Zulliger \textit{et al.} & 2022 & \cite{zulliger_within_2022} & Rest, MS & Inter, Intra & -- \\
65 & Allouch \textit{et al.} & 2023 & \cite{allouch_effect_2023} & Rest & Intra & Conn \\
66 & Borgheai \textit{et al.} & 2023 & \cite{borgheai_multimodal_2024} & -- & Intra & -- \\
67 & Dreyer \textit{et al.} & 2023 & \cite{dreyer_pauline_large_2023} & MI & Intra & Acc \\
68 & Ganin \textit{et al.} & 2023 & \cite{ganin_sources_2023} & ERP & Intra & -- \\
69 & García Alanis \textit{et al.} & 2023 & \cite{garcia_alanis_devils_2023} & -- & Inter, Intra & ICC \\
70 & Gupta \textit{et al.} & 2023 & \cite{gupta_raw_2023} & MI & Intra & Acc, Conn \\
71 & Huang \textit{et al.} & 2023 & \cite{huang_discrepancy_2023} & MI, Sleep & Inter, Intra, Sess & Acc, Shift \\
72 & Li \textit{et al.} & 2023 & \cite{li_eeg_2023} & -- & Intra & -- \\
73 & Nakuci \textit{et al.} & 2023 & \cite{nakuci_within-subject_2023} & Rest, Sleep & Inter, Intra, Sess & ICC, Conn \\
74 & Parameshwaran \& Thiagarajan & 2023 & \cite{parameshwaran_high_2023} & Rest & Intra & -- \\
75 & Saha \textit{et al.} & 2023 & \cite{saha_can_2023} & MI & Inter, Intra & Shift \\
76 & Sartzetaki \textit{et al.} & 2023 & \cite{sartzetaki_beyond_2023} & MI & Intra & Acc \\
77 & Shen \& Lin & 2023 & \cite{shen_cross-day_2024} & -- & Inter, Intra & -- \\
78 & Zhang \& Luck & 2023 & \cite{zhang_variations_2023} & ERP & Inter, Intra & ICC, SNR \\
79 & Ahuis \textit{et al.} & 2024 & \cite{ahuis_evaluation_2024} & -- & Intra, Data & SNR \\
80 & Chen \textit{et al.} & 2024 & \cite{chen_cross-subject_2024} & -- & Inter, Intra & Conn \\
81 & Dong \textit{et al.} & 2024 & \cite{dong_modeling_2024} & ERP & Inter, Intra, Trial & SNR, VarD \\
82 & Eyamu \textit{et al.} & 2024 & \cite{eyamu_prefrontal_2024} & ERP & Intra & -- \\
83 & Goswami \textit{et al.} & 2024 & \cite{goswami_improving_2024} & MI & Intra & Shift \\
84 & Jun \textit{et al.} & 2024 & \cite{jun_cognitive_2024} & Rest & Intra & Conn \\
85 & Liu \textit{et al.} & 2024 & \cite{liu_resting_2024} & Rest & Intra & ICC \\
86 & Meghdadi \textit{et al.} & 2024 & \cite{meghdadi_inter_2024} & Rest & Inter, Intra & ICC \\
87 & Memar \textit{et al.} & 2024 & \cite{memar_rise-ieeg_2024} & -- & Inter, Intra & -- \\
\end{longtable}
}

\clearpage
\section*{Appendix B: Selected studies by paradigm}
\label{app:paradigm}

\noindent
The studies below are those discussed by name in the paradigms section,
with the design, headline finding and stated limitation of each. It
complements Appendix~A, which codes all 87 included studies but carries no
narrative.

{\tiny
\setlength{\tabcolsep}{4pt}
\begin{longtable}{p{2.1cm} p{5.2cm} p{3cm} p{2.5cm} p{3cm}}
\caption{Variability Studies Across Major EEG Paradigms}
\label{tab:section4_variability}\\
\toprule
\textbf{Category} & \textbf{Study} & \textbf{Focus / Design} & \textbf{Key Finding} & \textbf{Limitation} \\
\midrule
\endfirsthead
\toprule
\textbf{Category} & \textbf{Study} & \textbf{Focus / Design} & \textbf{Key Finding} & \textbf{Limitation} \\
\midrule
\endhead
\midrule
\multicolumn{5}{r}{\emph{continued on next page}}\\
\endfoot
\bottomrule
\endlastfoot
\textbf{1. Resting-State} 
& Wang et al.~\cite{wang_test-retest_2022}
& 60 subjects, 3 sessions (90min + 30 days)
& State control critical; alpha ICC\textgreater
0.8
& Young adults; short-term only \\
\addlinespace
& Meghdadi et al.~\cite{meghdadi_inter_2024}
& Inter/intra frequency-specific ICC
& 7-9Hz most reliable (ICC\textgreater
0.94)
& Clinical sample limits generalization \\
\addlinespace
& Arazi et al.~\cite{arazi_magnitude_2016}
& Trial-by-trial reproducibility
& Trait-like across time/tasks
& Small sample; specific tasks \\
\addlinespace
& Nakuci et al.~\cite{nakuci_within-subject_2023}
& 8 sessions, multi-modal
& Within>between reproducibility; alpha stable
& Complex setup; small N \\
\addlinespace
& Scrivener \& Reader~\cite{scrivener_variability_2022}
& Between-subject electrode placement
& 3.94-7.17mm variability across axes
& No test-retest design \\
\midrule

\textbf{2. ERPs} 
& Zhang \& Luck~\cite{zhang_variations_2023}
& 40 participants, component quality
& ERP quality varies by component/paradigm
& Cross-sectional; no longitudinal \\
\addlinespace
& Ganin et al.~\cite{ganin_sources_2023}
& 19 subjects, P300 latency
& Latency variability impacts BCI
& Small N; single paradigm \\
\addlinespace
& Changoluisa et al.~\cite{changoluisa_electrode_2018}
& Cross-subject/session electrode selection
& Electrode choice affects P300 stability
& P300-specific; limited scope \\
\addlinespace
& Dong et al.~\cite{dong_modeling_2024}
& Population heterogeneity modeling
& Trial-level variability in auditory ERPs
& Auditory modality only \\
\addlinespace
& Bland \& Schaefer~\cite{bland_exploiting_2011}
& Trial-to-trial mechanisms
& Intra-recording variability structured
& No test-retest component \\
\midrule

\textbf{3. MI-BCI} 
& Saha \& Baumert~\cite{saha_intra-_2020}
& Review: inter/intra variability
& Both dimensions are major BCI challenges
& Review; no new empirical data \\
\addlinespace
& Huang et al.~\cite{huang_discrepancy_2023}
& Exp1 (inter) vs Exp2 (intra) comparison
& Cross-subject $\neq$ cross-session patterns
& N=10; single MI paradigm \\
\addlinespace
& Ma et al.~\cite{ma_large_2022}
& 25 subjects, 5 days, adaptation
& Accuracy: 68.8 $\rightarrow$ 53.7\%; adapt $\rightarrow$ 78.9\%
& 2-3 days interval; short-term \\
\addlinespace
& Zhou et al.~\cite{zhou_relative_2021}
& 20 subjects, 7 sessions, alpha power
& Alpha predicts MI across subjects/sessions
& Limited time period; single center \\
\addlinespace
& Borgheai et al.~\cite{borgheai_multimodal_2024}
& fNIRS + EEG multimodal prediction
& Predicts inter/intra performance ($R^2 = 0.942$)
& Requires dual modality; complexity \\
\midrule

\textbf{4. Advanced} 
& Huang et al.~\cite{huang_m3_2022}
& M$^3$CV: 106 subjects, 95 with 2 sessions
& Multi-paradigm (6) biometric stability
& Max 2 sessions; longer-term unknown \\
\addlinespace
& Melnik et al.~\cite{melnik_systems_2017}
& Systems-Subjects-Sessions variance
& Subjects=32\%, Systems=9\%, Sessions=1\%
& Small N=4; limited systems \\
\addlinespace
& Liu et al.~\cite{liu_reliability_2020}
& 54 subjects, 2-day microstate reliability
& Spatial stable; temporal moderate ICC
& 2-day only; short-term \\
\addlinespace
& Zulliger et al.~\cite{zulliger_within_2022}
& Within vs between-subject alpha-behavior
& Associations differ by analysis level
& Alpha-specific; single paradigm \\
\addlinespace
& García Alanis et al.~\cite{garcia_alanis_devils_2023}
& Multi-dimensional variability
& Characterizes cognitive EEG variability
& Complex; challenging interpretation \\
\end{longtable}
}

\bibliography{refs}

\end{document}